\documentclass[number,review]{elsarticle}

\usepackage{lineno,hyperref}
\modulolinenumbers[5]

\journal{Elsevier Journal}

\usepackage[dvipsnames]{xcolor}
\definecolor{sPointColor}{HTML}{d300ff}
\definecolor{pPointColor}{HTML}{ff0081}
\usepackage{graphicx}
\usepackage{amsmath,bm,amsfonts,mathtools,amssymb}
\usepackage[toc,page]{appendix}
\usepackage{array}
\usepackage{paralist}
\usepackage{booktabs}
\usepackage{multirow}
\usepackage{multicol}
\usepackage{tabularx}
\newcolumntype{Y}{>{\centering\arraybackslash}X}
\usepackage{subcaption}
\usepackage{ulem}
\usepackage[outline]{contour}
\usepackage[T1]{fontenc}
\usepackage{pifont}
\usepackage{longtable}
\usepackage{tabularray}

\usepackage{listofitems}

\usepackage{changepage}

\usepackage{pgfplotstable}
\usepackage{pgfplots}
\pgfplotsset{compat=1.9}
\usetikzlibrary{plotmarks}
\usetikzlibrary{positioning}
\usetikzlibrary{matrix}
\usetikzlibrary{external}
\usetikzlibrary{calc}
\usetikzlibrary{arrows}
\usetikzlibrary{decorations.markings}
\usetikzlibrary{shapes.geometric}
\usetikzlibrary{patterns}
\usepgfplotslibrary{fillbetween}
\usepgfplotslibrary{statistics}

\pgfplotsset{select coords between index/.style 2 args={
    x filter/.code={
        \ifnum\coordindex<#1\fi
        \ifnum\coordindex>#2\fi
    }
}}

\usepackage{algorithm}
\usepackage{algpseudocode}

\makeatletter

\newcommand{\norm}[1]{\left\lVert#1\right\rVert}

\newcommand{\tUnit}{\mathrm{T}}                                       %time
\newcommand{\aUnit}{\mathrm{L\,T^{-2}}}                               %acceleration
\newcommand{\lUnit}{\mathrm{L}}                                       %length
\newcommand{\pUnit}{\mathrm{M\,L^{-1}T^{-2}}}                         %pressure
\newcommand{\kPUnit}{\mathrm{L^{2}T^{-2}}}                            %kinematic pressure
\newcommand{\ndUnit}{\mathrm{-}}                                      %dimensionless
\newcommand{\lenC}{\ell_\mathrm{C}}                                    %converging part length
\newcommand{\lenD}{\ell_\mathrm{D}}                                    %diffuser length
\newcommand{\eeff}{e_{\mathrm{eff}}}

\newcommand{\pSet}{\mathcal{P}^*}                                      %pareto optimal set
\newcommand{\pFront}{\mathcal{O}^*}                                    %pareto front
\newcommand{\revs}[1]{\textcolor{black}{#1}}

\algnewcommand\algorithmicforeach{\textbf{for each}}
\algdef{S}[FOR]{ForEach}[1]{\algorithmicforeach\ #1\ \algorithmicdo}

\tikzset{cAlg/.style={ellipse,draw,minimum height=0.5cm,minimum width=0.8cm}}
\tikzset{cData/.style={ellipse,minimum height=0.5cm,minimum width=0.8cm,fill=gray,draw=gray,text=black,fill opacity=0.2,text opacity=1.0}}
\tikzset{cFunc/.style={rectangle,draw,minimum height=0.5cm,minimum width=0.8cm, align = center}}
\tikzset{cInfo/.style={rectangle,minimum height=0.5cm,minimum width=0.8cm,fill=gray,draw=gray,text=black,fill opacity=0.2,text opacity=1.0}}

\makeatother
\hypersetup{
    colorlinks=true,       % false: boxed links; true: colored links
    linkcolor=black,          % color of internal links
    citecolor=black,        % color of links to bibliography
    filecolor=black,      % color of file links
    urlcolor=black,           % color of external links
    allcolors=black
}

\begin{document}

\begin{frontmatter}

\title
{
    Accelerating shape optimization by deep neural networks with on-the-fly determined architecture
}

\address[ITCAS]
{
 Institute of Thermomechanics of the Czech Academy of Sciences,
 Dolej\v{s}kova 5, Prague 182~00, Czech Republic
}

\address[IHCAS]
{
 Institute of Hydrodynamics of the Czech Academy of Sciences,
 Pod Pa\v{t}ankou 5, Prague 166~12, Czech Republic
}

\address[VSCHTDM]
{
 University of~Chemistry and Technology, Prague,
 Department of~Mathematics, Informatics and Cybernetics,
 Technick\'{a}~5, Prague 166~28, Czech Republic
}

\address[VSCHT]
{
 University of~Chemistry and Technology, Prague,
 Department of~Chemical Engineering,
 Technick\'{a}~5, Prague 166~28, Czech Republic
}

\author[ITCAS,VSCHTDM]{Lucie Kub\'{i}\v{c}kov\'{a}}
\author[IHCAS,VSCHT]{Ond\v{r}ej Gebousk\'{y}}
\author[IHCAS]{Jan Haidl}
\author[ITCAS,VSCHTDM]{Martin Isoz\corref{cor}}
\cortext[cor]{Corresponding author,
tel: \mbox{+420 26605 2832}.}
\ead{isozm@it.cas.cz}
\ead[url]{https://www.it.cas.cz/en/d4/l041/}

\begin{abstract}
In component shape optimization, the component properties are often evaluated by computationally expensive simulations. Such optimization becomes unfeasible when it is focused on a global search requiring thousands of simulations to be evaluated. Here, we present a viable global shape optimization methodology based on multi-objective evolutionary algorithms accelerated by deep neural networks (DNNs). Our methodology alternates between evaluating simulations and utilizing the generated data to train DNNs with various architectures. When a suitable DNN architecture is identified, the DNN replaces the simulation in the rest of the global search. Our methodology was tested on \revs{a number of standardized benchmark functions}, showing itself at the level of and sometimes more flexible than other state-of-the-art acceleration approaches. Then, it was applied to a real-life optimization problem, namely the shape optimization of a single-phase ejector. Compared with a non-accelerated methodology, ours was able to save weeks of CPU time in solving this problem. \revs{Additionally, four optimized ejector shapes were 3D printed and tested experimentally to show their real-life performance.}

\end{abstract}


\end{frontmatter}

%~ \linenumbers

%%% INTRODUCTION %%%%%%%%%%%%%%%%%%%%%%%%%%%%%%%%%%%%%%%%%%%%%%%%%%%%%%%
\section{Introduction}
\label{sec:intro}
With ongoing technological advances, especially in 3D printing, device components of almost arbitrary shapes can be manufactured. Consequently, the design of the components can be tailored to a specific application, stimulating the development and utilization of various shape optimization methods. For components in contact with a fluid in motion, such as aircrafts, cars, or turbomachines, shape optimization methods are mostly backed by computational fluid dynamics (CFD) to evaluate component properties~\citep{skinner2018, li2017}. However, CFD simulations of real-life problems are costly, limiting the viability of CFD-based shape optimization methods~\citep{manriquez2016}.

One viable option to shape optimization is to use gradient-based methods (GBMs), such as the sparse nonlinear optimizer~\citep{gill2005} or the globally convergent method of moving asymptotes~\citep{svanberg1987}. These GBMs are well recognized for their computational efficiency, even when handling hundreds of design variables~\citep{foster1997}. When provided with a good enough starting point, they are well suited to find locally optimal solutions. However, they may struggle to find the global optimum~\citep{skinner2018}.

Another optimization approach builds on stochastic (gradient-free) methods (SMs) that are usually divided into three groups,
\begin{inparaenum}[(i)]
    \item particle swarm optimization~\citep{eberhart1995},
    \item evolutionary algorithms~\citep{goldbergBook1989, emmerich2018}, and
    \item simulated annealing~\citep{kirkpatrick1983}.
\end{inparaenum}
Contrary to the GBMs, none of these frameworks require continuity or predictability over the design space, and all of them are more resistant to getting trapped in a local optimum. However, as~\citet{yu2018} shows for the case of wing shape optimization, SMs usually require by an order of magnitude more function evaluations than GBMs. Consequently, numerous attempts have been made to combine SMs and GBMs to create hybrid methods that benefit from the strengths of both approaches, see, e.g.,~\citep{gage1995,vicini1999,kim2014}.

An alternative way to increase the viability of SMs is to pair them with computationally efficient surrogate models, creating the so-called surrogate-assisted optimization algorithms (SAOAs) ~\citep{manriquez2016}. The most widely used surrogate models are based on Gaussian process regression (aka Kriging models)~\citep{chugh2016, knowles2006}, radial basis functions~\citep{muller2017, dong2021} or deep neural networks~\citep{tian2023, wolday2024}. However, the choice and construction of a suitable surrogate model is often problem-dependent and may require extensive testing prior to the optimization itself, limiting the method generality and flexibility.

In this work, we propose a surrogate-assisted optimization algorithm that utilizes multi-objective evolutionary algorithms (MOEAs) paired with deep neural networks (DNNs) where the main novelty is that the most suitable DNN architecture is found during the optimization run. Specifically, the methodology cycles between running an optimization with a CFD-based cost function and training of DNNs on intermediate optimization results. In each cycle, DNNs with different architectures are tested, and the features of the best DNN are passed on to the next cycle. When a sufficiently capable DNN is found, it replaces CFD in the optimization, and CFD is used further only for verification and corrective means.

Our methodology is first benchmarked using the ZDT functions \revs{of}~\citet{zitzler2000} \revs{and LZ functions by~\citet{li2009} proposed} for testing multi-objective optimization algorithms. Its performance is compared with other similar state-of-the-art SAOAs~\citep{muller2017, morita2022, mastrippolito2021}. Next, the methodology is used to solve a real-life optimization problem; namely, the shape optimization of a single-phase ejector. The motivation for this problem choice was twofold. First, it represents a relatively simple task for CFD simulations. Second, we have access to a specifically tailored experimental setup that was eventually used to validate the optimization results. Furthermore, the performance of the ejector is known to strongly depend on its geometry, and the effects of alterations in the geometry of its various parts on its performance have been investigated in recent decades~\citep{havelka1997, utomo2009, daniels2019}. In our case, the shape optimization focused on the converging part and diffuser because these parts can be easily 3D printed and their optimized performance experimentally validated.

The paper proceeds as follows. \revs{In Section~\ref{sec:optimization}}, multi-objective optimization is introduced in general \revs{and} a description of multi-objective evolutionary algorithms \revs{is given}. \revs{In Section~\ref{sec:surrogates}}, we provide details on our surrogate-assisted optimization algorithm. The algorithm is based on an adaptive connection between DNNs and CFD and is hereafter abbreviated as CFDNNetAdapt. \revs{In Section~\ref{sec:verif}}, the benchmarking using ZDT functions is introduced\revs{, with further benchmarks based on LZ functions provided in Supplementary material.} Lastly, \revs{in Section~\ref{sec:application}}, the real-life optimization problem is described in detail, and the optimization results are discussed. Details on the experimental setup and the used CFD model are given in the appendix.

%%% OPTIMIZATION PROBLEM %%%%%%%%%%%%%%%%%%%%%%%%%%%%%%%%%%%%%%%%%%%%%%%
\section{Multi-objective shape optimization with CFD}
\label{sec:optimization}

\subsection{Multi-objective optimization problem}
\label{sub:mop}
Following the approach of~\citet{deb2001}, a multi-objective optimization problem is defined as
\begin{equation}
    \label{eq:moop}
    \begin{array}{*1{>{\displaystyle}c}}
    \min \bm{f}_\mathrm{cost} (\bm{p}) \\ [0.1cm]
    \text{subject\ to \ }\bm{c} \leq 0\text{\ and \ }\bm{p}_\mathrm{min} \leq \bm{p} \leq \bm{p}_\mathrm{max} \\ [0.1cm]
    \end{array}
\end{equation}
where $\bm{f}_\mathrm{cost}(\bm{p}) = \bm{o}$ is a cost function, $\bm{c}$ are considered constraints and $\bm{p}_\mathrm{min}$ the minimum and $\bm{p}_\mathrm{max}$ the maximum parameter bounds.

In case the objectives are anti-correlated, i.e., improving performance in one objective may worsen performance in the others, there does not exist a single optimal solution. Instead, a set of trade-off solutions is sought. This set is commonly referred to as the Pareto-optimal set ($\pSet$). Its counterpart with trade-off objective values is the Pareto-optimal front ($\pFront$)~\citep{pareto1896}.

Trade-off solutions are identified as non-dominated solutions in terms of Pareto-dominance~\citep{pareto1896}. A non-dominated solution is defined as a solution that is not dominated by any other available solution. One solution ($\bm{p}_1$) dominates another ($\bm{p}_2$) if
\begin{equation}
    \label{eq:dominance}
    \begin{array}{*1{>{\displaystyle}c}}
    f_{\mathrm{cost},i}(\bm{p}_1) \leq f_{\mathrm{cost},i}(\bm{p}_2)\ \text{ for all}\  i,\ i = 1,\,\dots,\,m \\ [0.1cm]
    \text{and} \\ [0.1cm]
    f_{\mathrm{cost},i}(\bm{p}_1) < f_{\mathrm{cost},i}(\bm{p}_2)\ \text{ for at least one\ }i,\ i = 1,\,\dots,\,m \\ [0.1cm]
    \end{array}
\end{equation}
where $m$ is the number of objectives.

\subsection{Multi-objective evolutionary algorithms}
\label{sub:moea}
To solve a multi-objective optimization problem, we opted for using the multi-objective evolutionary algorithms (MOEAs)~\citep{emmerich2018}. Of the many MOEA variants available, we used here the NSGA-II algorithm by~\citet{deb2002}. The algorithm works with a population of individua\revs{ls} that is updated each iteration. Each individu\revs{al} is a tuple $(\bm{p},\bm{f}_\mathrm{cost}(\bm{p}))$. The population size ($n_\mathrm{pop}$) is preset and the first population consists of individua\revs{ls} with randomly generated vectors $\bm{p}$ within a range defined by $\bm{p}_\mathrm{min}$ and $\bm{p}_\mathrm{max}$.

In each iteration, random pairs of individua\revs{ls} are chosen and mixed using partially random operators such as crossover and mutation, generating offspring individua\revs{ls} with different $\bm{p}$ vectors. The offspring individua\revs{ls} are included in the population, and all the individua\revs{ls} are sorted. In NSGA-II, the sorting is based on the Pareto-dominance principle. Lastly, the population size is reduced to $n_\mathrm{pop}$ by truncating the least fit individua\revs{ls} identified by the sorting~\citep{deb2002}. After this truncation, a new iteration starts. Note that each iteration is also referred to as a generation. The algorithm terminates after a preset number of generations ($n_\mathrm{gen}$).

\subsection{Baseline optimization framework}
\label{sub:cfdoptfram}
\begin{figure}[htbp]
    \centering
    %~ \input{\myGraphs/baselineFrameworkV1.tex}
    \includegraphics{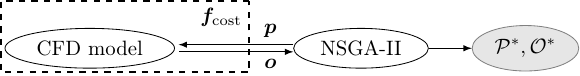}
    \caption{Schematic of the baseline optimization framework. The vectors $\bm{p}$ and $\bm{o}$ contain the parameters and objectives, respectively. Furthermore, $\bm{f}_\mathrm{cost}$ is the cost function, $\pSet$ the Pareto-optimal set and $\pFront$ the Pareto-optimal front.}
    \label{fig:baselineFramework}
\end{figure}

To have a baseline, a not-surrogate-assisted framework was prepared. The framework schematic is given in Figure~\ref{fig:baselineFramework}.

In this work, the baseline framework is used to generate
\begin{inparaenum}[(i)]
    \item{a referential solution (resulting $\pSet,\,\pFront$) for comparison with surrogate-assisted optimization algorithms, and}
    \item{data points (all evaluated tuples ($\bm{p}$,$\bm{o}$)) for training of surrogate models.}
\end{inparaenum}

%%% SURROGATE AIDED OPTIMIZAION ALGORITHM %%%%%%%%%%%%%%%%%%%%%%%%%%%%%%
\section{Surrogate-assisted optimization}
\label{sec:surrogates}

The bottleneck of CFD-driven optimizations is usually the long evaluation time of CFD simulations. Therefore, surrogate-assisted optimization algorithms (SAOAs) are often used to speed up the optimization. We prepared such a framework utilizing deep neural networks, namely multilayer perceptrons (MLPs).

Our proposed framework is similar to the adaptive evolution control frameworks as detailed in a review by~\citet{manriquez2016}. A novelty with respect to there-listed SAOAs lies in the fact that our framework (CFDNNetAdapt) includes an adaptive on-the-fly determination of a suitable MLP architecture. Note that even though our framework was designed for shape optimization having CFD as the expensive-to-evaluate cost function, it can be applied to any optimization problem with real-valued cost function.

\subsection{Multilayer perceptrons}
\label{sec:dnns}
A multilayer perceptron (MLP) is a feed-forward neural network that comprises several layers of neurons that are fully connected~\citep{goodfellow2016}. The first is the input layer, then there are several hidden layers, and the last is the output layer. In our framework, the MLP takes the parameters $\bm{p}$ as input and provides a prediction of the objectives $\bm{o}$ as the output. Therefore, the size of the first and last layer is equal to the number of parameters and objectives, respectively. The number of hidden layers is fixed (hyperparameter $n_\mathrm{hlrs}$) and their size is optimized on-the-fly.

\begin{figure}[htbp]
    \centering
    \includegraphics{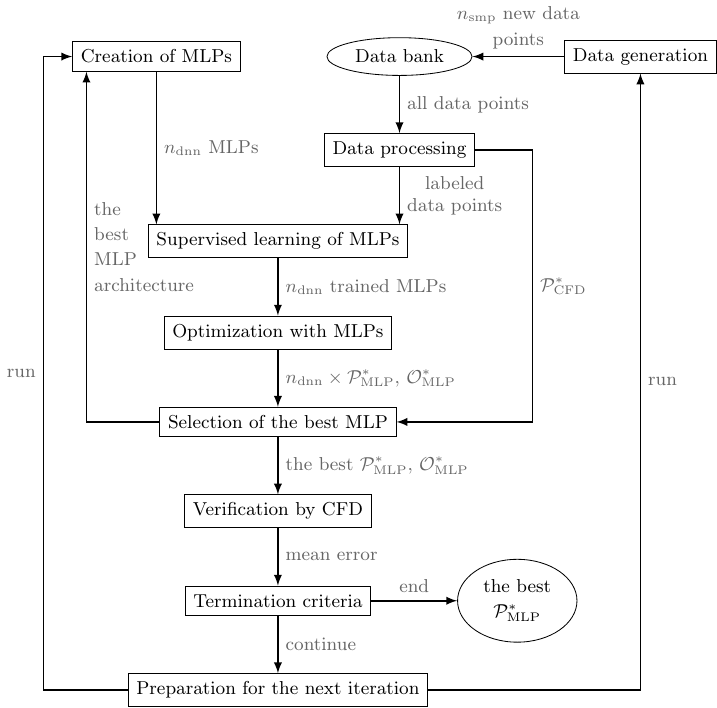}
    \caption{Schematics of steps and data-sharing in CFDNNetAdapt. Used hyperparameters are listed in Table~\ref{tab:dnnhypers}. MLP stands for "multilayer perceptron". Grey indicates passed data or signals.} %~ Boxes are cross-linked with paragraphs in the text.}
    \label{fig:cfdnnetadapt}
\end{figure}

\begin{table}
    \centering
    \small
    \renewcommand{\arraystretch}{0.8}
    \revs{
    \begin{tabular}{clc}
        \toprule
        hyperparameter & description \\
        \midrule
        $n_\mathrm{hlrs}$ & number of hidden layers \\
        $n_\mathrm{dnn}$ & number of MLPs generated in each iteration \\
        $r_n$ & half width of the interval for choosing hidden layer size \\
        $n_\mathrm{max}$ & maximal size of MLP layer \\
        $n_\mathrm{min}$ & minimal size of MLP layer \\
        $\bar{n}^0_i$ & initial value for the average of the $i$-th hidden layer size \\
        $n_\mathrm{smp}$ & number of data points generated in one iteration \\
        $n_\mathrm{pop}$ & population size for NSGA-II \\
        $n_\mathrm{gen}$ & number of generations for NSGA-II \\
        $n_\mathrm{ver}$ & number of points chosen for verification \\
        $\epsilon$ & required tolerance \\
        $k_\mathrm{max}$ & maximum number of iterations \\
        \bottomrule
    \end{tabular}
    }
    \renewcommand{\arraystretch}{1}
    \caption{List of CFDNNetAdapt hyperparameters with description. The hyperparameters are ordered in the same way as they are mentioned in text.}
    \label{tab:dnnhypers}
\end{table}

\subsection{CFDNNetAdapt}
\label{sec:cfdnnetadapt}
The schematic of CFDNNetAdapt is given in Figure~\ref{fig:cfdnnetadapt}. Its individual steps are described in detail in the following paragraphs. In addition, the hyperparameters of CFDNNetAdapt are summarized in Table~\ref{tab:dnnhypers}.

\paragraph{Creation of MLPs}
\label{par:creation}
CFDNNetAdapt operates in iterations; in each iteration, a number of MLPs (hyperparameter $n_\mathrm{dnn}$) is created. The transfer function used was always \textit{tanh}. The number of hidden layers (hyperparameter $n_\mathrm{hlrs}$) is the same for all MLPs, but the size of the hidden layers is varied. For the $i$-th hidden layer, its size is chosen randomly from an interval defined by an average value ($\bar{n}_i$) and half of its width (hyperparameter $r_n$), i.e., 
\begin{equation}
    \label{eq:random}
    n_i = \min(n_\mathrm{max},\, \max(n_\mathrm{min},\,\tilde{n}_i)),\quad \tilde{n}_i = \text{random}(\bar{n}_i - r_n,\,\bar{n}_i + r_n)\,,
\end{equation}
where $n_\mathrm{max}$ and $n_\mathrm{min}$ are hyperparameters defining the maximal and minimal allowed layer size, respectively. Note that the interval half-width is kept the same for all the hidden layers, but the average value can be different. In the first iteration, the averages are preset (hyperparameters $\bar{n}^0_i$, $i=1,\,\dots,\,n_\mathrm{hlrs}$). In subsequent iterations, the averages are inherited from the best MLP found in the previous iteration.

\paragraph{Data generation}
\label{par:generation}
To perform a supervised learning of the created MLPs, data points (tuples ($\bm{p},\bm{o}$)) are required. A preset number of data points (hyperparameter $n_\mathrm{smp}$) is generated by the baseline optimization framework described in Section~\ref{sub:cfdoptfram} and saved in a data bank. If data points from previous iterations are present, the baseline framework is not started from scratch, but selects its initial population from the data bank.

\paragraph{Data processing}
\label{par:processing}
All data stored in the data bank are processed together. First, non-dominated points are identified and saved as an intermediate Pareto-optimal set, $\pSet_\mathrm{CFD}$. Next, all data points are \revs{sorted} into three groups for supervised learning of MLPs. Specifically, \revs{the data points are randomly shuffled} and \revs{the first} $75\,\%$ are labeled as training data, \revs{the next} $15\,\%$ as validation data, and \revs{the last} $10\,\%$ as testing data.

\paragraph{Supervised learning of MLPs}
\label{par:learning}
Each MLP is trained on the training data using the Levenberg-Marquardt algorithm~\citep{levenberg1944, marquardt1963}. \revs{To prevent overfitting, the validation data are used to monitor the MLP performance during training~\citep{goodfellow2016} and the training termination is decided based on non-successful training iterations. In particular, after each training iteration, the MLP is used to evaluate the validation data points, and the mean error between the predicted and true values is computed. If the error is greater than the error from the previous iteration, the MLP receives one strike. If the total number of strikes becomes greater than a threshold value (value of $10\cdot n_\mathrm{hlrs}$ was used in this work), the training is terminated.}

After the training is completed, testing data are used to evaluate the performance of each MLP, and the results are saved to allow for during-run or after-run checks of the algorithm progress. Inclusion of the performance check based on the testing data is motivated by an effort to allow the user to identify the potential overfitting of adaptively selected MLP architectures to the validation data.

\paragraph{Optimization with MLPs}
\label{par:optimization}
\revs{With each trained MLP, a separate optimization is run where the MLP acts as a surrogate model in cost function evaluation.} %~ Each trained MLP is used as a surrogate in an optimization run by NSGA-II.
\revs{The optimizations are run by NSGA-II} with a predefined population size (hyperparameter $n_\mathrm{pop}$) and number of generations (hyperparameter $n_\mathrm{gen}$). Hence, each MLP predicts its own Pareto-optimal set ($\pSet_\mathrm{MLP}$) and front ($\pFront_\mathrm{MLP}$).

\paragraph{Selection of the best MLP}
\label{par:selection}
\revs{From the $n_\mathrm{dnn}$ trained MLPs, the best one has to be selected. This selection is based on the similarity of the MLP-predicted Pareto optimal sets ($\pSet_\mathrm{MLP}$) to the intermediate Pareto-optimal set ($\pSet_\mathrm{CFD}$) identified from the data points. The MLP that predicted the most similar Pareto-optimal set is chosen as the best one.}

\revs{The similarity of the Pareto-optimal sets is evaluated using the inverted generational distance (IGD) defined in agreement with~\citep{li2009} as}
\begin{equation}
    \label{eq:igd}
    \text{IGD}(\pSet_\mathrm{CFD},\,\pSet_\mathrm{MLP}) = \frac{1}{\|\pSet_\mathrm{CFD}\|}\displaystyle\sum_{\bm{x} \in \pSet_\mathrm{CFD}} \min_{\bm{y} \in \pSet_\mathrm{MLP}} \norm{\bm{x},\,\bm{y}}\,\revs{.}
\end{equation}
\revs{The Pareto-optimal set $\pSet_\mathrm{MLP}$ with the lowest IGD value is perceived as the most similar, and the corresponding MLP is chosen as the best MLP.}

This choice, which may be seen as \revs{a method of} under-relaxation, originates from the decreasing reliability of sets with a higher IGD. When a poorly trained MLP is used in an optimization, it tends to exploit gaps in its training and come up with solutions that are far off from what it was trained on, thus potentially having a large prediction error. Therefore, the Pareto-optimal set with the lowest IGD is assumed to contain the lowest number of such exploited solutions.

The MLP-predicted Pareto-optimal set with the lowest IGD, the corresponding Pareto-optimal front, and the MLP that was used to create these are saved. The rest of the sets, fronts, and MLPs are discarded.

\paragraph{Verification by CFD}
\label{par:verification}
The suitability of the most promising MLP identified based on IGD is verified by CFD. A number (hyperparameter $n_\mathrm{ver}$) of solutions is randomly chosen from the MLP-predicted Pareto-optimal set ($\pSet_\mathrm{MLP}$). For each chosen solution, a corresponding CFD model is constructed and evaluated. It is possible that some of the chosen points cannot be evaluated by CFD, i.e., the computational mesh would be too non-orthogonal or skewed. In such cases, CFDNNetAdapt randomly chooses substitute solutions from the set. 

For the evaluated solutions, the predicted and re-evaluated objective values are compared, and the  mean error is computed as
\begin{equation}
    \label{eq:delta}
    \delta := \displaystyle\frac{1}{n_{\mathrm{ver}}}\displaystyle\sum\limits_{i=1}^{n_{\mathrm{ver}}} \norm{\bm{o}_{\mathrm{MLP},i} - \bm{o}_{\mathrm{CFD},i}}
\end{equation}
where $\bm{o}_\mathrm{MLP}$ and $\bm{o}_\mathrm{CFD}$ are vectors of predicted and true objectives, respectively.

\paragraph{Termination criteria}
\label{par:termination}
If the verification error is smaller than a predefined tolerance ($\epsilon$), the CFDNNetAdapt run is terminated. The most suitable Pareto-optimal set ($\pSet_\mathrm{MLP}$) and the currently best-performing MLP are returned. Another reason for termination is reaching a preset maximum number of iterations of CFDNNetAdapt (hyperparameter $k_\mathrm{max}$).

\paragraph{Preparation for the next iteration}
\label{par:preparation}
In the event that the algorithm did not end, the architecture of the most suitable MLP found in the current iteration is forwarded to the next iteration. In particular, the hidden layer sizes of the best MLP are used as the average sizes in the creation of new MLPs.

\paragraph{\revs{Note on stochastic nature of CFDNNetAdapt}}
\revs{First, CFDNNetAdapt uses a MOEA (in our case NSGA-II) as a data generator. Thus, it inherently includes stochastic operations. Second, beyond the standard sources of randomness in MOEAs, CFDNNetAdapt also introduces randomness through variations in the sizes of the hidden layers of the MLPs used as surrogates. The random selection of hidden-layer sizes, followed by evaluating each MLP performance via IGD is motivated by the need to identify MLPs that have good generalization capabilities and are not prone to overfitting. However, this procedure adds an additional layer of stochasticity to the behavior of CFDNNetAdapt. To illustrate the effects of this randomness, further details and a statistical analysis of the variability in MLP architectures across all optimization problems considered in this work are provided in \ref{sec:heuristics}.}

\subsection{\revs{Implementation details}}
\label{sec:implementation}

\revs{To conclude the description of the CFDNNetAdapt algorithm, we briefly describe implementations of (i) the baseline optimization framework, and (ii) CFDNNetAdapt. In addition, CFDNNetAdapt is an open-source project, and all source codes and configuration files necessary to reproduce the computations presented in this study are available from \url{https://github.com/techMathGroup/CFDNNetAdapt}.}

\revs{Specifically, both the baseline and the surrogate-aided frameworks are implemented in Python and prepared to run on high-performance computing architectures. The MOEAs are included through an open-source Python library for multi-objective evolutionary optimization \textit{Platypus}~\citep{platypus}. Everything related to MLPs is implemented using an open-source Python library \textit{pyrenn}~\citep{pyrenn}. These libraries are combined with a custom code for the evaluation of $f_\mathrm{cost}$. Moreover, for the purposes of the real-life application, the $f_\mathrm{cost}$ evaluation comprises an automated preparation of a CFD model in OpenFOAM~\citep{OpenFOAM2007}. The description of the CFD model construction for our real-life application (optimization of an ejector pump) is given in~\ref{sec:mathMod}.}

%%% BENCHMARKING %%%%%%%%%%%%%%%%%%%%%%%%%%%%%%%%%%%%%%%%%%%%%%%%%%%%%%%
\section{Benchmarking of CFDNNetAdapt}
\label{sec:verif}
For the verification of CFDNNetAdapt, \revs{several benchmark optimization problems were chosen including the ZDT functions} suggested by~\citet{zitzler2000} \revs{and LZ functions proposed by~\citet{li2009}. In this section, the results of verification tests using the ZDT functions are presented. The results obtained using the LZ functions are similar in nature to the results  obtained using ZDT functions. Therefore, the verification results from the LZ functions are presented in the Supplementary material.}

\revs{For the verification presented here, we used five of the ZDT functions which} are referred to as ZDT1, ZDT2, ZDT3, ZDT4, and ZDT6. All of these operate with real parameters and two objectives. The test function ZDT5 was not included as its parameters are decoded as bitstrings. The complete specification of the test functions chosen is given in Table~\ref{tab:zdts}. Optimal solutions to each problem are illustrated in black in Figure~\ref{fig:zdtsObj}.

\begin{table}
    \centering
    \small
    \begin{adjustwidth}{-1.0cm}{-1.0cm}
    \begin{tabularx}{1.2\textwidth}{p{0.12\textwidth}cYc}
        \toprule
        name & parameters & cost functions & optimal sol. \\ [0.2cm]
        \midrule
             & $n = 10$ & $f_1(\bm{p}) = p_1$ & $p_1 \in [0,1]$ \\ [0.1cm]
        ZDT1 & $p_i \in [0,1]$ & $f_2(\bm{p}) = g(\bm{p})\,\left[1 - \sqrt{p_1/g(\bm{p})}\right]$ & $p_i = 0$ \\ [0.1cm]
             & $i = 1,\,\dots,\,n$ & $g(\bm{p}) = 1 + 9\,(\sum^n_{i=2} p_i)/(n - 1)$ & $i = 2,\,\dots,\,n$ \\ [0.1cm]
        \midrule
             & $n = 10$ & $f_1(\bm{p}) = p_1$ & $p_1 \in [0,1]$ \\ [0.1cm]
        ZDT2 & $p_i \in [0,1]$ & $f_2(\bm{p}) = g(\bm{p})\,\left[1 - (p_1/g(\bm{p}))^2\right]$ & $p_i = 0$ \\ [0.1cm]
             & $i = 1,\,\dots,\,n$ & $g(\bm{p}) = 1 + 9\,(\sum^n_{i=2} p_i)/(n - 1)$ & $i = 2,\,\dots,\,n$ \\ [0.1cm]
        \midrule
             & $n = 10$ & $f_1(\bm{p}) = p_1$ & $p_1 \in [0,1]$ \\ [0.1cm]
        ZDT3 & $p_i \in [0,1]$ & $f_2(\bm{p}) = g(\bm{p})\,\left[1 - \sqrt{p_1/g(\bm{p})} - \frac{p_1}{g(\bm{p})}\,\mathrm{sin}(10\pi p_1)\right]$ & $p_i = 0$ \\ [0.1cm]
             & $i = 1,\,\dots,\,n$ & $g(\bm{p}) = 1 + 9\,(\sum^n_{i=2} p_i)/(n - 1)$ & $i = 2,\,\dots,\,n$ \\ [0.1cm]
        \midrule
             & $n = 2$ & $f_1(\bm{p}) = p_1$ & $p_1 \in [0,1]$ \\ [0.1cm]
        ZDT4 & $p_1 \in [0,1]$ & $f_2(\bm{p}) = g(\bm{p})\,\left[1 - \sqrt{p_1/g(\bm{p})}\right]$ & $p_i = 0$ \\ [0.1cm]
             & $p_2 \in [-5,5]$ & $g(\bm{p}) = 1 + 10(n-1) + \sum^n_{i=2}\left[p^2_i - 10\,\mathrm{cos}(4\pi p_i)\right]$ & $i = 2,\,\dots,\,n$ \\ [0.1cm]
        \midrule
             & $n = 2$ & $f_1(\bm{p}) = 1 - \mathrm{exp}(-4p_1)\,\mathrm{sin}^6\,(6\pi p_1)$ & $p_1 \in [0,1]$ \\ [0.1cm]
        ZDT6 & $p_i \in [0,1]$ & $f_2(\bm{p}) = g(\bm{p})\left[1 - (f_1(\bm{p})/g(\bm{p}))^2\right]$ & $p_i = 0$ \\ [0.1cm]
             & $i = 1,\,\dots,\,n$ & $g(\bm{p}) = 1 + 9\left[(\sum^2_{i = 2}p_i)/(n-1)\right]^{0.25}$ & $i = 2,\,\dots,\,n$ \\ [0.1cm]
        \bottomrule
    \end{tabularx}
    \end{adjustwidth}
    \caption{Utilized ZDT functions~\citep{zitzler2000}.}
    \label{tab:zdts}
\end{table}

\begin{table}[htbp]
    \centering
    \revs{
    \begin{tabular}{cc|cc|cc}
        hyperpar. & value & hyperpar. & value & hyperpar. & value \\
        \midrule
        $n_\mathrm{hlrs}$ & $3$ & $n_\mathrm{min}$ & $2$ & $n_\mathrm{gen}$ & $250$ \\
        $n_\mathrm{dnn}$ & $4$ & $n_\mathrm{max}$ & $20$ & $n_\mathrm{ver}$ & $16$ \\
        $\{\bar{n}^0_1,\,\bar{n}^0_2,\,\bar{n}^0_3\}$ & $\{11,\,11,\,11\}$ & $n_\mathrm{smp}$ & $1,000$ & $k_\mathrm{max}$ & $100$ \\
        $r_n$ & $4$ & $n_\mathrm{pop}$ & $100$ & $\epsilon$ & $10^{-6}$
    \end{tabular}
    }
    \caption{Values of CFDNNetAdapt \revs{hyper}parameters used for the ZDT functions. \revs{Individual hyperparameters are explained in Table~\ref{tab:dnnhypers}.}}
    \label{tab:setzdts}
\end{table}

\subsection{\revs{Compared algorithms and algorithm settings}}
The performance of CFDNNetAdapt was compared with two widely used algorithms
\begin{inparaenum}[(i)]
    \item{NSGA-II without any surrogate model~\citep{deb2002}, and }
    \item{state-of-the-art surrogate-assisted optimization algorithm SOCEMO~\citep{muller2017}.}
\end{inparaenum}
In addition, a variant of CFDNNetAdapt with non-DNN surrogate model was constructed to investigate the suitability of DNN usage. \revs{As the non-DNN surrogate model, we have chosen models based on Gaussian process regression (GPR)~\citep{jones1998}. The choice of GPR was motivated by its frequent use (mainly as Kriging models) in SAOAs, as described in, e.g.~\citep{morita2022,mastrippolito2021}. Hereafter, this variant will be referred to as CFDGPRAdapt.} A more detailed description of each method settings is provided in the following paragraphs, and the results are depicted in Figures~\ref{fig:zdtsObj}--~\ref{fig:zdtsInd}.

\revs{Similarly to~\citet{deb2002}, NSGA-II was run with population size $n_\mathrm{pop} = 250$ and number of generations $n_\mathrm{gen} = 100$. All evaluated data points (tuples ($\bm{p}$,$\bm{o}$)) generated by NSGA-II ($\sim \!\! 25,000$) were used to feed the other SAOAs. The SAOAs training sets were gradually extended as follows: the first batch comprised $10$ samples, the second $30$, then  $100; 300; 500;$ and $1,000; 2,000;\dots;25,000$ samples.}

\revs{The setting of CFDNNetAdapt hyperparameters is given in Table~\ref{tab:setzdts} and was kept the same for all the ZDT functions to have the comparison fair. To set the hyperparameters defining the MLP size ($n_\mathrm{hlrs}$, $n_\mathrm{max}$), a few MLPs with small to large architectures (increasing both $n_\mathrm{hlrs}$ and $n_\mathrm{max}$) were trained using from $200$ to $1,000$ NSGA-II samples generated for ZDT1. After the training, a mean error of the MLPs was evaluated on unseen NSGA-II samples. For $n_\mathrm{hlrs} \geq 3$ and $n_\mathrm{max} \geq 20$, the error stagnated at $\delta_{\mathrm{stg}} \sim 10^{-6}$. Based on this testing, the $n_\mathrm{hlrs}$ and $n_\mathrm{max}$ were set as the limit values. Also, the $\epsilon$ could be set as $\epsilon  \sim \delta_{\mathrm{stg}} \approx 10^{-6}$ and $n_\mathrm{ver}$ was estimated to provide a trade-off between accuracy and computational costs.}

\revs{Next, the hyperparameter $n_\mathrm{min}$ was set as the smallest possible value, i.e., $n_\mathrm{min} = 2$, and the values of $\{\bar{n}_1,\,\bar{n}_2,\,\bar{n}_3\}$ and $r_n$ were computed as}
\begin{equation}
    \label{eq:nirnhypers}
    \bar{n}^0_i = \frac{1}{2}\left(n_\mathrm{max} + n_\mathrm{min}\right),\quad r_n = \frac{1}{4}\left(n_\mathrm{max} - n_\mathrm{min}\right),\,
\end{equation}
\revs{Lastly, the values of $n_\mathrm{pop}$ and $n_\mathrm{gen}$ were set the same as for pure NSGA-II. Based on the total number of NSGA-II samples ($\sim 25,000$), the $n_\mathrm{smp}$ was chosen as $1,000$ and the expected number of iterations is then $25$. The maximum number of iterations $k_\mathrm{max}$ was chosen as four times the expected number.} 

\revs{Next,} the surrogates used by SOCEMO are based on radial basis functions (RBFs) and were originally used for problems with up to $500$ data points. As we work with up to $25,000$ points, we implemented a probabilistic selection of inducing points based on~\citep{uhrenholt2021} to effectively lower the problem dimension.

\revs{Finally, the setting of CFDGPRAdapt was kept as close as possible to the setting of CFDNNetAdapt presented in Table~\ref{tab:setzdts}. Only the settings used for MLPs could not be used and new settings for GPRs was added.} Specifically, the sparse variant of the GPR with up to $50$ inducing points was used. In each iteration, four GPR models with different kernels (squared exponential, Matern52, Matern32, and Matern12) were tested. The implementation was done using a Python library \textit{gpflow}~\citep{gpflow2017}.  

\subsection{\revs{Estimates of Pareto-optimal solutions}}
\begin{figure}[htbp]
    \centering
    \includegraphics{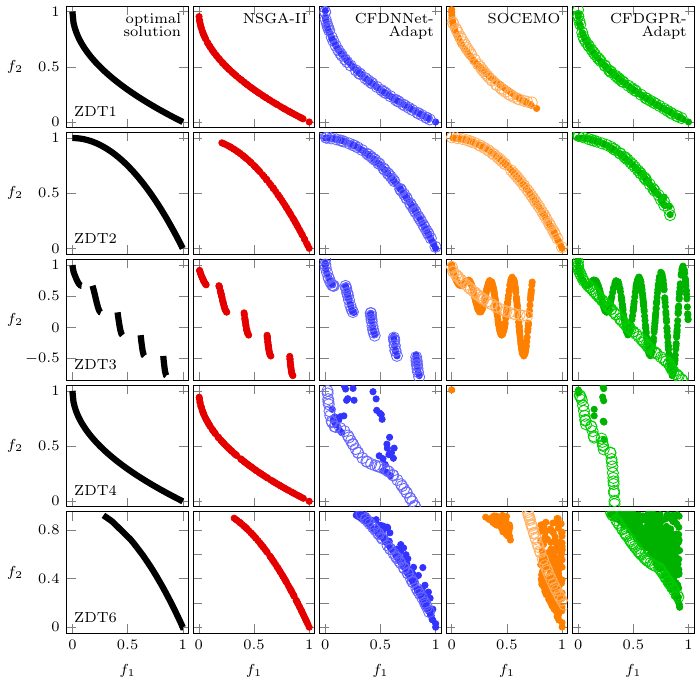}
    \caption{Comparison of \revs{Pareto-optimal front estimates for} the ZDT problems. From left to right, there are optimal solutions (black), solutions found by NSGA-II (red), solutions found by CFDNNetAdapt (blue), solutions found by SOCEMO (orange), and solutions found by \revs{CFDGPRAdapt} (green). The empty circles depict the Pareto-optimal front as predicted by the surrogate model, the full circles show the corresponding true values.}
    \label{fig:zdtsObj}
\end{figure}

\revs{In Figure~\ref{fig:zdtsObj}, for each SAOA,} two estimates of Pareto-optimal fronts corresponding to the same Pareto-optimal set are depicted: the one predicted by the best found surrogate model (empty circles) and its counter-part re-evaluated by the true cost functions (full circles). \revs{The solutions (Pareto-optimal fronts) found by NSGA-II are shown in red. CFDNNetAdapt solutions are depicted in blue. The solutions of SOCEMO~\citep{muller2017} are shown in orange. Lastly, the solutions found by \revs{CFDGPRAdapt} are depicted in green.}

\revs{With $25,000$ function evaluations, NSGA-II converged to a good estimate of the Pareto-optimal front for all the tested functions. As a result, high-quality training samples for SAOAs were always available. On the other hand, for ZDT4 and all the $25,000$ samples used for training, none of the SAOAs was able to converge to the Pareto-optimal solution. For the other ZDT functions, CFDNNetAdapt consistently outperformed the other tested SAOAs.}

\begin{figure}[htbp]
    \centering
    \includegraphics{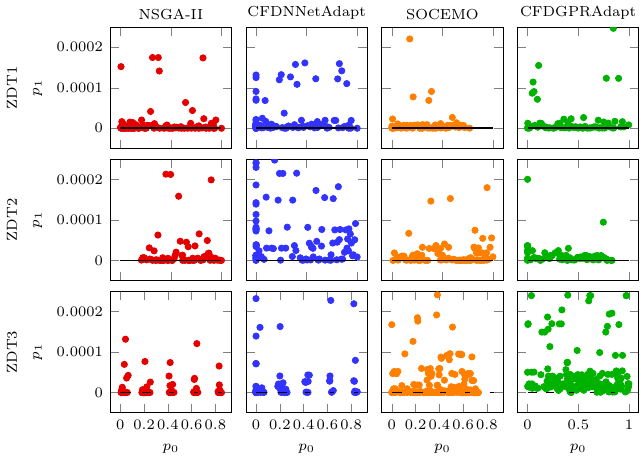}
    \caption{\revs{Comparison of Pareto-optimal set estimates for ZDT problems and all the available data used for SAOAs training. Pareto-optimal sets are shown in black. Next, from left to right, there are solutions found by NSGA-II (red), solutions found by CFDNNetAdapt (blue), solutions found by SOCEMO (orange), and solutions found by {CFDGPRAdapt} (green).}}
    \label{fig:zdtsParAll}
\end{figure}
\revs{In Figure~\ref{fig:zdtsParAll} are shown the estimates of the Pareto-optimal sets corresponding to the results in Figure~\ref{fig:zdtsObj} for which at least one SAOA converged. Here, note that for ZDT1, both CFDNNetAdapt and CFDGPRAdapt were able to estimate the Pareto-optimal set similarly to NSGA-II while SOCEMO was not able to identify the solutions with $p_0 \gtrsim 0.8$. On the other hand, CFDNNetAdapt solutions for ZDT2 are plagued by a number of parasitic points. Still, for ZDT3, CFDNNetAdapt was the only SAOA able to correctly identify individual groups of Pareto-optimal solutions.}

\revs{Similar results to that shown in Figures~\ref{fig:zdtsObj} and~\ref{fig:zdtsParAll} but for the LZ functions are provided in Supplementary material. The results for LZ functions confirm the capability of CFDNNetAdapt to identify even complex-shaped Pareto-optimal sets; see Figures~7 and~8 in Supplementary material.}

\subsection{\revs{Evaluating algorithm convergence}}
\begin{figure}[htbp]
    \centering
    \includegraphics{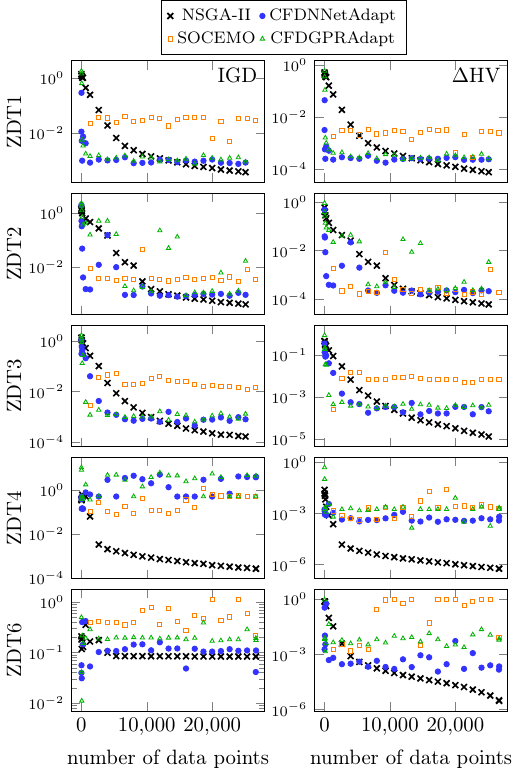}
    \caption{Evolution of the IGD~\eqref{eq:igd} and $\Delta$HV~\eqref{eq:hvzdts} indicators from NSGA-II, CFDNNetAdapt, SOCEMO, and \revs{CFDGPRAdapt}.}
    \label{fig:zdtsInd}
\end{figure}
In addition to the comparison of Pareto-optimal fronts in Figure~\ref{fig:zdtsObj}, the convergence of the mentioned algorithms was also examined. For each algorithm, the evolution of two indicators is depicted in Figure~\ref{fig:zdtsInd}. The first indicator is the inverted generational distance defined in \eqref{eq:igd}. Each resulting Pareto-optimal set $\mathcal{P}^*$ was compared to the optimal set $\mathcal{P}^*_\mathrm{opt}$, i.e., $\text{IGD}(\mathcal{P}^*_\mathrm{opt},\,\mathcal{P}^*)$ was computed.

The second indicator was based on the hyper-volume (HV) indicator. In two-dimensional objective space, HV represents an area that is non-dominated by points of a given Pareto-optimal front ($\pFront$) with respect to a preset reference point ($\bm{r}$). \revs{In agreement with~\citep{guerreiro2021}, the hypervolume indicator can be defined as} 
\begin{equation}
    \label{eq:hv}
    \revs{
    \text{HV}(\pFront, \bm{r}) = \Lambda\left(\displaystyle\bigcup_{\substack{\bm{p}\in\pFront \\[0.1cm]{\bm{p}} \text{\ dominates\ }\bm{r}}}\left[\bm{p},\bm{r}\right]\right),\,
    }
\end{equation}
\revs{where $\Lambda(\cdot)$ is the Lebesgue measure and $\left[\bm{p},\bm{r}\right]$ is a box defined by the points $\bm{p}$ and $\bm{r}$. For each optimization problem, the reference point was chosen as $\bm{r} = \left[\displaystyle\max_{\text{NSGA-II}}(f_1),\displaystyle\max_{\text{NSGA-II}}(f_2)\right]$. For a given $\pFront$ and $\bm{r}$, the dimension-sweep algorithm~\citep{fonseca2006} was used to compute the $\text{HV}$ indicator.}

For our comparison, the scaled difference of HV to optimal HV ($\Delta \text{HV}$) was examined. It was computed as
\begin{equation}
    \label{eq:hvzdts}
    \Delta \text{HV}(\pFront_\mathrm{opt}, \pFront, \bm{r}) = \frac{\text{HV}(\pFront_\mathrm{opt},\bm{r}) - \text{HV}(\pFront,\bm{r})}{\text{HV}(\min(\pFront_\mathrm{opt}),\bm{r})}
\end{equation}
where $\pFront_\mathrm{opt}$ comprises the optimal objective values\revs{, and} $\text{HV}(\min(\pFront_\mathrm{opt}),\bm{r})$ represents the area of a rectangle defined by the minimal optimal values of the cost functions and the reference point.

\subsection{\revs{Discussion of algorithm performance}}
For the problems ZDT1 and ZDT2, the SAOAs mostly converged to lower indicator values faster than NSGA-II. However, limitations are visible in Figure~\ref{fig:zdtsObj} for ZDT1-SOCEMO and ZDT2-GPRs where the algorithms were unable to uncover the whole Pareto-optimal front. This is also apparent in the convergence graphs in Figure~\ref{fig:zdtsInd}, where in the respective cases the indicators jump by orders of magnitude even in the latter iterations. In one iteration, the surrogates are able to locate the whole front, in the subsequent they are not. In CFDNNetAdapt, the information about the best surrogate architectures is passed along iterations, which presumably helps to avoid this problem.

In ZDT3, the SAOAs still converged faster, but the problems were larger. Surrogate models based on GPRs and RBFs had considerable problems with the discontinuity of the Pareto-optimal front. CFDNNetAdapt appears to be the most flexible. In ZDT4, all SAOAs failed. None of the surrogate models was able to give accurate enough predictions and most of the solutions predicted to lie on the Pareto-optimal front (empty circles in Figure~\ref{fig:zdtsObj}) were actually far off the true values (full circles in Figure~\ref{fig:zdtsObj}).

Lastly, in ZDT6, the inaccuracy of the predictions is also apparent. Nevertheless, the algorithms were able to locate the Pareto-optimal front. The least smudged front was provided by CFDNNetAdapt that also achieved the lowest indicator values. Moreover, the IGD indicator shows that even NSGA-II had problems identifying the whole Pareto-optimal set, the indicator stuck at about $10^{-1}$.

Overall, CFDNNetAdapt works sufficiently well in comparison with other available options. The usage of DNNs appears to be adequate when compared with models based on GPRs or RBFs. In addition, the flexibility of DNNs together with the CFDNNetAdapt feature of passing architecture-relevant information to successive iterations gives the algorithm a slight advantage when dealing with non-continuous or complex optimization search spaces.

Still, the standardized tests performed suggest that, similarly to other SAOAs, the CFDNNetAdapt algorithm is not universally usable. The main strength of CFDNNetAdapt lies in accelerating optimizations with computationally intensive evaluation of cost functions. Based on examining Figures~\ref{fig:zdtsObj} and~\ref{fig:zdtsInd}, a best-practice approach is proposed. First, CFDNNetAdapt is the most advantageous for optimizations in topologically simple parameter and objective spaces (similar to ZDT1 and ZDT2). Second, for cases with a complex objective space but a simple parameter space (ZDT3), it can still provide satisfactory solutions with significantly fewer data points than NSGA-II. Third, for the cases of topologically complex parameter space (ZDT4 and ZDT6), the algorithm will fail. This failure is to be expected because in such cases, small variances in cost function parameters significantly affect the returned objectives, which makes the construction of any data-driven surrogate problematic at best.

\begin{table}
    \centering
    \small
    \revs{
    \begin{adjustwidth}{-1.0cm}{-1.0cm}
    \begin{tabularx}{1.2\textwidth}{ccccc}
        \toprule
        function & algorithm & mean IGD (std. dev.) & mean $\Delta$HV (std. dev.) & function evaluations \\
        \midrule
             & CFDNNetAdapt & 9.03E-04 (9.99E-05) & 2.42E-04 (2.26E-05) & [14; 13,778] \\
        ZDT1 & SOCEMO       & 2.18E-02 (9.71E-03) & 1.78E-03 (8.61E-04) & [1,333; 4,443] \\
             & CFDGPRAdapt  & 1.14E-03 (1.24E-04) & 2.91E-04 (4.50E-05) & [54; 12,445] \\
        \midrule
             & CFDNNetAdapt & 9.10E-04 (9.82E-05) & 2.05E-04 (2.34E-05) & [23; 13,334] \\
        ZDT2 & SOCEMO       & 2.44E-02 (6.90E-02) & 3.05E-03 (8.42E-03) & [1,333; 9,331] \\
             & CFDGPRAdapt  & 3.85E-02 (9.07E-02) & 4.83E-03 (1.09E-02) & [89; 1,823] \\
        \midrule
             & CFDNNetAdapt & 7.84E-04 (1.96E-04) & 2.71E-04 (9.54E-05) & [14; 11,556] \\
        ZDT3 & SOCEMO       & 1.41E-02 (3.95E-03) & 6.09E-03 (1.63E-03) & [1,333; 4,443] \\
             & CFDGPRAdapt  & 1.34E-03 (3.80E-04) & 4.42E-04 (1.24E-04) & [27; 9,778] \\
        \midrule
             & CFDNNetAdapt & 2.60E+00 (1.55E+00) & 4.38E-04 (1.46E-04) & [89; 667] \\
        ZDT4 & SOCEMO       & 8.35E-01 (1.06E+00) & 9.66E-03 (2.30E-02) & [N/A; N/A] \\
             & CFDGPRAdapt  & 2.29E+00 (1.75E+00) & 8.22E-04 (8.09E-04) & [134; 667] \\
        \midrule
             & CFDNNetAdapt & 1.22E-01 (2.81E-02) & 3.71E-04 (3.28E-04) & [14; 178] \\
        ZDT6 & SOCEMO       & 5.36E-01 (2.76E-01) & 5.60E-01 (4.02E-01) & [N/A; N/A] \\
             & CFDGPRAdapt  & 2.50E-01 (8.49E-02) & 1.39E-02 (1.25E-02) & [27; 223] \\
        \bottomrule
    \end{tabularx}
    \end{adjustwidth}
    }
    \caption{\revs{Numerical characteristics of the IGD~\eqref{eq:igd} and $\Delta$HV~\eqref{eq:hvzdts} indicators for CFDNNetAdapt, SOCEMO and CFDGPRAdapt. For each indicator, its mean value and standard deviation are given. Moreover, a range of dataset sizes for which the given SAOA is cost-effective is provided in the last column ($[\overline{\text{feval}}_\text{min};\underline{\text{feval}}_\text{max}]$). N/A marks the runs for which SAOA did not outperform the NSGA-II.}}
    \label{tab:zdtsstats}
\end{table}

\revs{Finally, in a manner similar to~\citep{deb2014,chugh2016,zhu2024}, in Table~\ref{tab:zdtsstats}, we provide numerical characteristics of the behavior of the three algorithms tested (CFDNNetAdapt, SOCEMO, and CFDGPRAdapt). In particular, for each algorithm, we provide the characteristic values of the mean inverse generational distance (mean IGD) and the difference in the hyper-volume indicator (mean $\Delta$HV) to which the algorithms converge with the increasing number of data points. The mean values were computed from the IGD and $\Delta$HV obtained for five separate algorithm runs and for the five largest datasets tested. The mean values of both IGD and $\Delta$HV are complemented by their standard deviations.}

\revs{Furthermore, to allow quantification of the computational costs of the individual algorithms, Table~\ref{tab:zdtsstats} contains data on the number of function evaluations in which the given algorithm outperformed NSGA-II in both IGD and $\Delta$HV for the first time $\overline{\text{feval}}_\text{min}$ and for the last time $\underline{\text{feval}}_\text{max}$. The value of $\overline{\text{feval}}_\text{min}$ is computed as the maximum over both IGD and $\Delta$HV and five separate algorithm runs, while the value $\underline{\text{feval}}_\text{max}$ is computed similarly, but as a minimum.}

\revs{The values $\overline{\text{feval}}_\text{min}$ and $\underline{\text{feval}}_\text{max}$ are defined in such a manner that they represent the minimum and maximum size of the dataset for which the tested SAOA is, under the condition that the cost function evaluation is significantly more resource-demanding than the surrogate construction, computationally cheaper than the baseline NSGA-II. Results in Table~\ref{tab:zdtsstats} suggest, that CFDNNetAdapt consistently provides a wider interval $[\overline{\text{feval}}_\text{min};\underline{\text{feval}}_\text{max}]$ than both SOCEMO and CFDGPRAdapt. Also, when comparing the other two indicators, IGD and $\Delta$HV, CFDNNetAdapt at least matches the performance of the remaining two algorithms.}

\begin{figure}[htbp]
    \centering
    \includegraphics{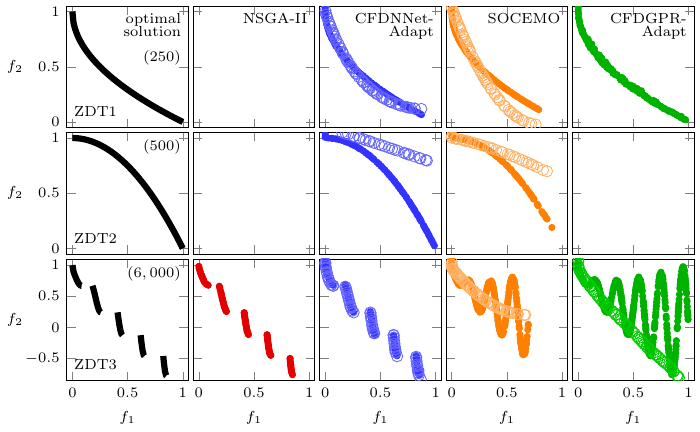}
    \caption{\revs{Comparison of solutions to the ZDT problems and reduced training datasets. Sizes of training datasets are given in parentheses in the first column of the figure. For description of colors and symbols, see Figure~\ref{fig:zdtsObj}.}}
    \label{fig:zdtsObjXXX}
\end{figure}
\begin{figure}[htbp]
    \centering
    \includegraphics{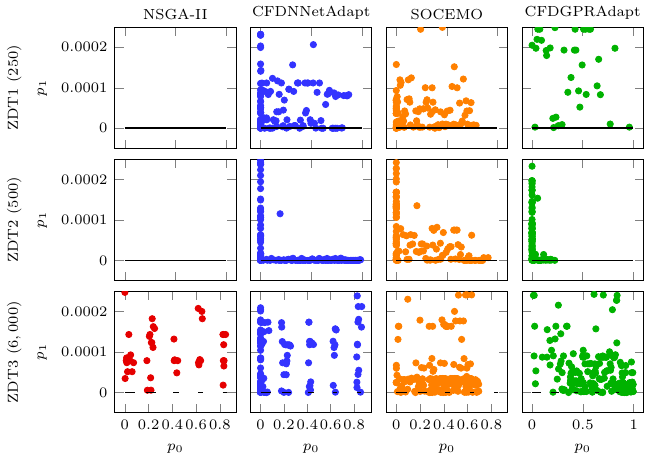}
    \caption{\revs{Comparison of Pareto-optimal set estimates for ZDT problems and reduced training datasets. Sizes of training datasets are given in parentheses next to function names. For description of colors and symbols, see Figure~\ref{fig:zdtsParAll}.}}
    \label{fig:zdtsParXXX}
\end{figure}
\revs{Still, for ZDT1 and ZDT3, based on the data in Table~\ref{tab:zdtsstats} and Figure~\ref{fig:zdtsInd}, CFDNNetAdapt appears to perform almost identically to CFDGPRAdapt. Similarly, for ZDT2, CFDNNetAdapt seems on par with SOCEMO. To clarify these inconclusive cases, the ZDT1, ZDT2, and ZDT3 functions were analyzed similarly as shown in Figures~\ref{fig:zdtsObj} and~\ref{fig:zdtsParAll}. However, this time, only a reduced number of datapoints was used for training of the surrogate-aided optimization algorithms. The size of the training dataset varied between the three tested functions but was kept the same for all the three tested SAOAs. Furthermore, the number of training samples  reflected the complexity of the optimization task, and was selected based on the evolution of IGD and $\Delta$HV shown in Figure~\ref{fig:zdtsInd}. Consequently, $250$ NSGA-II-generated samples were used for ZDT1, and $500$ and $6,000$ samples were used for ZDT2 and ZDT3, respectively.}

\revs{The estimates of Pareto-optimal fronts and Pareto-optimal sets obtained using this reduced number of tranining samples are shown in Figures~\ref{fig:zdtsObjXXX} and~\ref{fig:zdtsParXXX}. For ZDT1 and 250 training samples, both CFDNNetAdapt and SOCEMO were able to provide acceptable estimates of Pareto-optimal set while incorrectly placing the Pareto-optimal front. Such a situation occurs when the surrogate model can correctly estimate the trends in the mapping between parameter and objective space but the mapping itself is burdened by an error. The third algorithm, CFDGPRAdapt, provided the worst estimate of the Pareto-optimal front and also its estimate of the Pareto-optimal set is of lower quality compared to both CFDNNetAdapt and SOCEMO. However, comparing the number of required function evaluations and the quality of the $\pFront$ and $\pSet$ estimates, all the three SAOAs significantly outperformed pure NSGA-II.}

\revs{For ZDT2, the situation was similar to the one observed for ZDT1. Still, this time only CFDNNetAdapt was able to provide a suitable estimate of $\pSet$. For ZDT3, a significantly higher number of function evaluations was required for any of the SAOAs to be able to provide an acceptable $\pFront$ and $\pSet$ estimates. Furthermore, in this case only CFDNNetAdapt was able to outperform NSGA-II. Specifically, the provided estimates of $\pFront$ shown in Figure~\ref{fig:zdtsObjXXX} were similar to each other. However, as can be seen in the last row of Figure~\ref{fig:zdtsParXXX}, CFDNNetAdapt was able to provide a better estimate of $\pSet$ than NSGA-II.}

\revs{To conclude, in the cases where surrogate-aided optimization can be applied, CFDNNetAdapt usually provides better trade-off between the quality of approximation of the Pareto-optimal solution and the cost function evaluations than all the other tested approaches. In particular, for ZDT functions and approximately $100$ data points available for training, it significantly outperforms NSGA-II and this effectivity is maintained up to $\approx 2,500$ datapoints. Then, using more than $2,500$ data points is beneficial to ensure the CFDNNetAdapt convergence but the advantages over NSGA-II are less apparent. Finally, at $\approx 10,000$ data points, CFDNNetAdapt is usually outperformed by pure NSGA-II. Still, the ranges of required cost function evaluations for which CFDNNetAdapt outperforms NSGA-II are consistently wider than for both SOCEMO and CFDGPRAdapt. The superior performance of CFDNNetAdapt compared to SOCEMO and CFDGPRAdapt was observed not only for ZDT functions, but also for LZ functions. A detailed analysis for LZ functions is available in Supplementary material.}

%%% REAL-LIFE APPLICATION %%%%%%%%%%%%%%%%%%%%%%%%%%%%%%%%%%%%%%%%%%%%%%
\section{Real-life application}
\label{sec:application}
To illustrate the behavior of the CFDNNetAdapt algorithm in a real-life application, we have selected the shape optimization of a single-phase ejector pump. Ejector pumps are technologically simple devices that utilize the kinetic energy of a high-speed fluid jet to entrain a secondary fluid. Given their simple structure and absence of moving parts, ejectors are widely used, e.g., for the transport of fuel in aircraft wings~\citep{marini1992}, for the compression of coolant in refrigeration systems~\citep{taleghami2018}, etc. In addition to its relevance for process engineering, the focus on an ejector in this work was motivated by the fact that we have access to a specifically designed experimental setup, for details see~\ref{sec:experiment}. The availability of a highly modifiable experimental setup was leveraged to validate the performance of the optimized shapes.

\begin{figure}[htbp]
	\centering
	\includegraphics{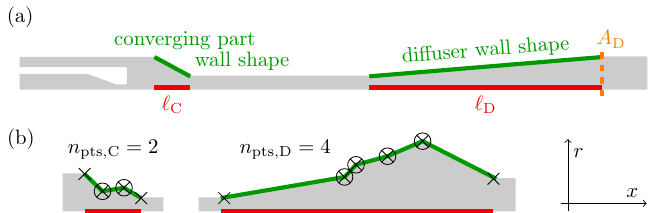}
	\caption{(a) Schematic of an ejector with the optimized parts highlighted in green and red. The plane $A_\mathrm{D}$ serves as a reference in the definition of ejector performance indicators. (b) Illustrative shapes of the converging part (left) and diffuser (right). The shapes are defined with fixed points ($\times$) and optimized control points ($\otimes$).}
	\label{fig:opPar}
\end{figure}
\subsection{Shape parametrization}
\label{sub:pars}
The optimization focused on the converging part and the diffuser of the ejector. The schematic of the ejector with the regions of interest highlighted is given in Figure~\ref{fig:opPar}a.
 
The length and shape of the outer wall of the parts were optimized. The wall shape was designed as piecewise linear. This approach was selected in order to promote the final design robustness with respect to the boundary layer detachment, especially at the diffuser start. Examples of possible piecewise designs are in Figure~\ref{fig:opPar}b.

Eventually, the design of the ejector was defined by two lengths ($\lenC$ and $\lenD$) and coordinates of $n_\mathrm{pts}$ control points ($n_\mathrm{pts} = n_\mathrm{pts,C} + n_\mathrm{pts,D}$) as indicated in Figure~\ref{fig:opPar}b. Given the axisymmetric geometry of the ejector, each control point is defined by two coordinates, hence the total number of parameters in the optimization was $n_\mathrm{pars} = 2\,n_\mathrm{pts} + 2$. We chose to have $n_\mathrm{pts,C} = 2$ control points that define the design of the converging part and $n_\mathrm{pts,D} = 4$ control points for the diffuser. Consequently, the total number of parameters was $n_\mathrm{pars} = 14$.

\subsection{Constraints}
\label{sub:cons}
Two design constraints were included. First, the total length of the experimental device is limited and therefore the sum of the lengths had to be kept below a predefined maximal value, i.e.
\begin{equation}
    \label{eq:lencon}
    \lenC + \lenD \leq \ell_\mathrm{max}\,.
\end{equation}

Second, the control points must be sequentially ordered along the $x$ axis. Having $n_\mathrm{pts}$ points, it must hold that
\begin{equation}
    \label{eq:ptscon}
    x_i > x_{i-1}\quad i = 2,\,\dots,\,n_\mathrm{pts}\,.
\end{equation}
The second constraint is crucial for effective automatic generation of meshes for CFD models.

\subsection{Optimized objectives}
\label{sub:opObj}
The geometry of the ejector was optimized subject to two performance indicators:
\begin{inparaenum}[(i)]
    \item the ejector energy efficiency ($\eeff$), and
    \item the total length ($\ell_\mathrm{t}$).
\end{inparaenum}
The first indicator reflects the primary purpose of an ejector being used as a pump. Thus, its energy efficiency is of utmost interest and is defined as
\begin{equation}
    \label{eq:eeff}
    \eeff = \frac{\overline{\dot{Q}}_{\mathrm{suction}} (\overline{p}_{\mathrm{D}} - \overline{p}_{\mathrm{suction}})}{\overline{\dot{Q}}_{\mathrm{inlet}} (\overline{p}_{\mathrm{inlet}} - \overline{p}_{\mathrm{D}})}\,,\quad
    \overline{\varphi}_A = \frac{1}{\norm{A}}\int\limits_{A}\varphi\,dS\,,
\end{equation}
where $\overline{\dot{Q}}$ and $\overline{p}$ are the surface-averaged volumetric flow rate and pressure, respectively. The subscripts in~\eqref{eq:eeff}$_1$ indicate the surfaces on which the averages are computed, i.e., \textit{inlet}, \textit{suction}, and the plane $A_{\mathrm{D}}$ as indicated in Figure~\ref{fig:opPar}a.

The energy efficiency was evaluated by CFD; particularly, using the OpenFOAM~\citep{OpenFOAM2007} library. Details about the generation of a CFD model, model numerical settings, and validation are given in \ref{sec:mathMod}. For each ejector design, the CFD model was evaluated for three primary fluid flow rates ($\dot{Q}_\mathrm{inlet} = \{0.3,\ 0.4,\ 0.5\}$~l/s) and an average value $\overline{e}_\mathrm{eff}$ was relevant for optimization.

In addition to $\eeff$, which is to be maximized, the second performance indicator, total length, reflects possible geometric constraints of the environment the ejector should be installed in; thus, it is to be minimized. The total length was defined as
\begin{equation}
    \label{eq:totl}
    \ell_\mathrm{t} = \lenC + \lenD\,,
\end{equation}
and according to our prior studies~\citep{kubickova2021}, minimization of $\ell_\mathrm{t}$ is expected to be in anti-correlation with maximization of $\eeff$.

\subsection{Penalization}
\label{sub:penalty}
In case the ejector geometry did not satisfy the design constraints \eqref{eq:lencon} and \eqref{eq:ptscon}, the performance indicators were not evaluated. Instead, a penalty was computed as
\begin{equation}
    \label{eq:penfun}
    s_\mathrm{cons} = 1.0 + \beta_\ell\cdot\left(\lenC + \lenD - \ell_\mathrm{max}\right) + \beta_\mathrm{pts}\cdot\displaystyle\sum_{\substack{i = 2 \\ x_i < x_{i-1}}}^{n_\mathrm{pts}} \left(x_{i-1} - x_i\right)\,,
\end{equation}
where $\beta_\ell$ and $\beta_\mathrm{pts}$ are constant weights that were set by trial and error as $\beta_\ell = 1.0$ and $\beta_\mathrm{pts} = 10.0$.

Moreover, even though the design satisfied the constraints, it could happen that the generated CFD mesh was too skewed and non-orthogonal and, therefore, unsuitable for CFD calculation. In such cases, the values of skewness ($\gamma_\mathrm{skew}$) and non-orthogonality ($\gamma_\text{non-o}$) indicators computed by OpenFOAM were used to compute a penalty as
\begin{equation}
    \label{eq:penm}
    s_\mathrm{mesh} = 1.0 + \beta_s\cdot\gamma_\mathrm{skew} + \beta_n\cdot\gamma_\text{non-o}
\end{equation}
where $\beta_\mathrm{skew}$ and $\beta_\text{non-o}$ are weights that were estimated by trial and error as $\beta_\mathrm{skew} = \beta_\text{non-o} = 0.1$.

\subsection{Cost function}
\label{sub:costfun}
In summary, the cost function distinguished three cases. First, the constraints \eqref{eq:lencon} and \eqref{eq:ptscon} were checked, and if either failed, the $s_\mathrm{cons}$ value was used. Second, the computational mesh was generated and its quality was estimated. When the quality of the mesh was insufficient, the value $s_\mathrm{mesh}$ was used. Third, the CFD model was calculated and the ejector performance indicators \eqref{eq:eeff} and \eqref{eq:totl} were evaluated.

The cost function ($\bm{f}_\mathrm{cost}$) can be written as
\begin{equation}
    \label{eq:costfun}
    \bm{o} = \bm{f}_\mathrm{cost}(\bm{p}) = \left\{
    \begin{array}{rl}
        s_{\mathrm{cons}}(\bm{p}) \cdot (1,1)^{\text{T}} & \text{in first case} \\ [0.2cm]
        s_{\mathrm{mesh}}(\bm{p}) \cdot (1,1)^{\text{T}} & \text{in second case} \\ [0.2cm]
        (-\overline{e}_\mathrm{eff}, \ell_\mathrm{t})^{\text{T}} & \text{in third case,}
    \end{array}
    \right.
\end{equation}
where $\bm{p}$ is the vector of parameters and $\bm{o}$ the vector of objectives. Note that the cost function was minimized and therefore the energy efficiency was multiplied by $-1$.

\subsection{Generation of common data points for comparison}
\label{sub:dimOptimRes}
The described optimization problem was first pre-solved using the baseline framework described in Section~\ref{sub:cfdoptfram}. The reason for this is to create a common set of data points for a fair comparison of CFDNNetAdapt with the baseline framework. In particular,
\begin{inparaenum}[(i)]
    \item CFDNNetAdapt will use the common data set as a source of data points instead of computing its own.
    \item The data points will be supplied to CFDNNetAdapt in the same order as they were generated by the baseline framework to mimic the conditions when CFDNNetAdapt creates its own data points.
    \item To compare CFDNNetAdapt-aided optimization with the baseline framework, the baseline framework will be restarted, and the common data set will be used as a starting point.
\end{inparaenum}

To generate the common data points, the NSGA-II algorithm was run with population size $n_\mathrm{pop} = 320$ and number of generations $n_\mathrm{gen} = 30$. Overall, $9,600$ ejector geometries were tested. However, approximately $25\,\%$ of the tested geometries could not be evaluated by CFD, since they either violated the design constraints or the computational mesh was too skewed and non-orthogonal, see Section~\ref{sub:penalty}. Therefore, the number of designs evaluated by CFD was about $7,200$. Each design was evaluated for three primary fluid flow rates and the total number of evaluated CFD models was $3 \cdot 7,200 = 21,600$.

\begin{figure}[htbp]
    \centering
    \includegraphics{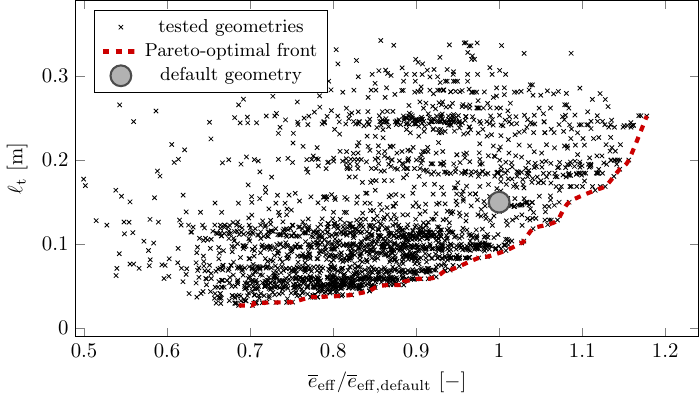}
    \caption{Results generated by the baseline framework in pre-solving of the optimization problem. Due to clarity, only a representative selection of the tested geometries, approximately a third of all the evaluated geometries, is depicted.}
    \label{fig:shapepof}
\end{figure}
The common data points together with the Pareto-optimal front found by the baseline framework are depicted in Figure~\ref{fig:shapepof}. For clarity of the graph, only a representative selection of all the ejector geometries is depicted. The selection was created such that no two ejector designs are too close to each other in the displayed objective space. In particular, when two points were closer than $10^{-2}$ in terms of Euclidean distance, one of them was omitted. 

The illustrated data show that the Pareto-optimal front is not equally resolved in all parts of the objective space, especially in parts with $\ell_\mathrm{t} > 0.1$~m and $\overline{e}_\mathrm{eff}/\overline{e}_\mathrm{eff,default} > 1.0$, the front is more sparse. Presumably, this is caused by the high dimensionality of the parameter space. Nonetheless, the Pareto-optimal front position did not change significantly in the last $5$ generations of the NSGA-II run.

\subsection{Comparison of CFDNNetAdapt and the baseline framework}
\label{sub:resdnn}
The common data points were used to supply a run of CFDNNetAdapt with data and as a starting point for the continuation of the baseline framework. Ideally, CFDNNetAdapt should find new ejector designs that will
\begin{inparaenum}[(i)]
    \item fill in the sparse regions of the pre-solved (common) Pareto-optimal front and
    \item be non-dominated by any of the already found solutions.
\end{inparaenum}
This should be achieved at a fraction of the computational costs required to continue the baseline framework until it converges to similarly good results.

\begin{table}[htbp]
    \centering
    \begin{tabular}{cc|cc|cc}
        hyperpar. & value & hyperpar. & value & hyperpar. & value \\
        \midrule
        $n_\mathrm{hlrs}$ & $3$ & $n_\mathrm{min}$ & $2$ & $n_\mathrm{gen}$ & $90$ \\
        $n_\mathrm{dnn}$ & $10$ & $n_\mathrm{max}$ & $20$ & $n_\mathrm{ver}$ & $8$ \\
        $\{\bar{n}^0_1,\,\bar{n}^0_2,\,\bar{n}^0_3\}$ & $\{11,\,11,\,11\}$ & $n_\mathrm{smp}$ & $600$ & $k_\mathrm{max}$ & $12$ \\
        $r_n$ & $4$ & $n_\mathrm{pop}$ & $500$ & $\epsilon$ & $0.05$
    \end{tabular}
    \caption{Values of CFDNNetAdapt parameters used for the real-life application.}
    \label{tab:setreal}
\end{table}
In each of its iterations, CFDNNetAdapt took $n_\mathrm{smp} = 600$ points from the common data set. The points were taken in the same order as they were generated by the baseline framework. Regarding the CFDNNetAdapt hyperparameters, \revs{they are listed in Table~\ref{tab:setreal}.} More details \revs{on} the hyperparameters \revs{selection} are given in \revs{Table~\ref{tab:dnnhypers} and its description}.

\begin{figure}[htbp]
    \centering
    \includegraphics[width=0.8\textwidth]{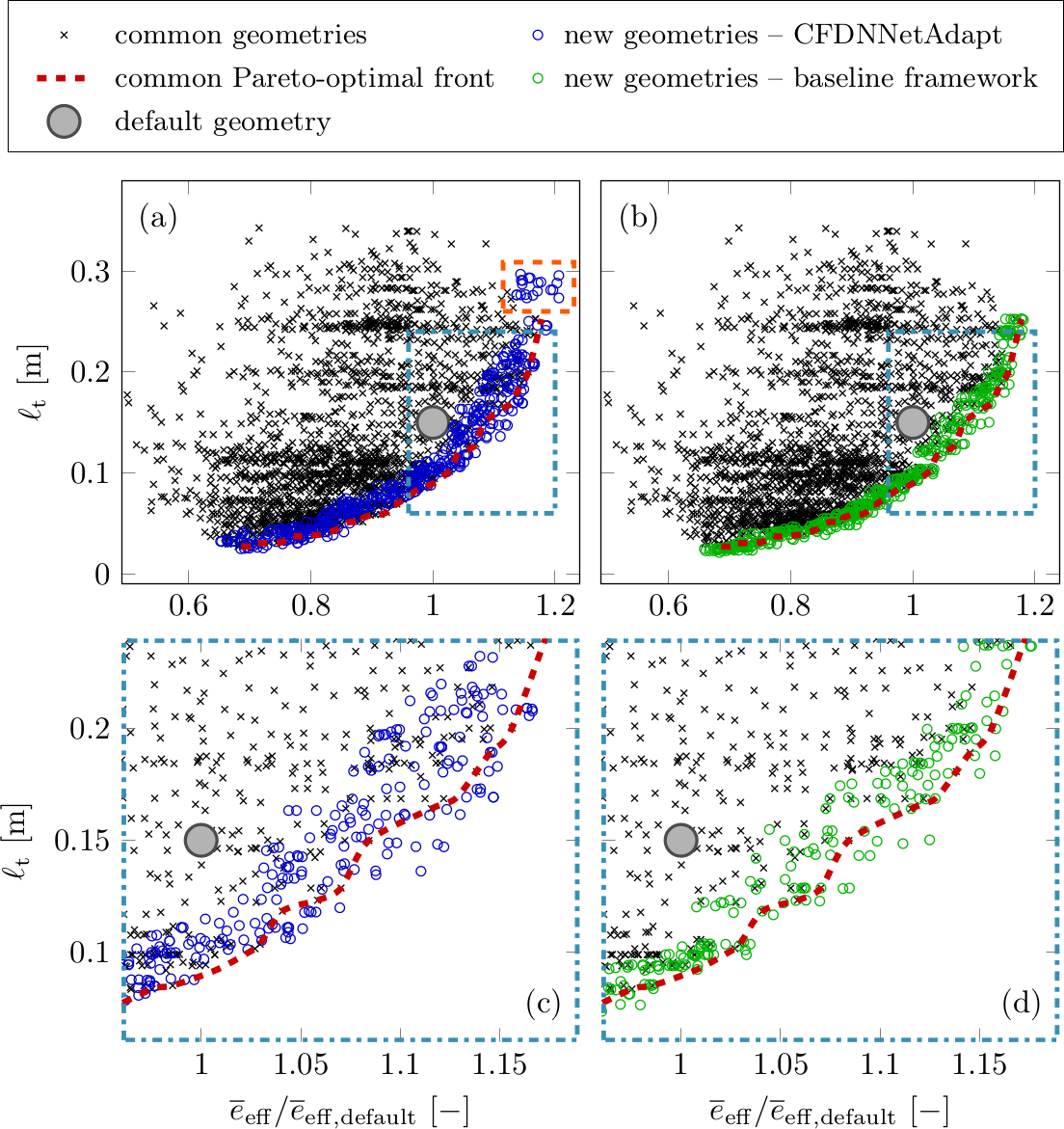}
    \caption{Common data points from Figure~\ref{fig:shapepof} displayed with new ejector geometries. (a) New geometries were found by CFDNNetAdapt with the values of $\overline{e}_\mathrm{eff}$ recomputed by CFD. (b) New geometries found by continuation of the baseline framework. (c) and (d) Zoom on the data in dash-dotted teal boxes in (a) and (b), respectively. Geometries found by CFDNNetAdapt only are highlighted by a dashed orange box.}
    \label{fig:annvscfd}
\end{figure}
It took CFDNNetAdapt 10 iterations ($6,600$ samples) to achieve verification error $\delta = 0.036$, see~\eqref{eq:delta}, and to meet the required tolerance $\epsilon = 0.05 \approx 5\,\%$ average error in the optimized objectives. The best MLP found had $5$ fully connected layers with $14,\,5,\,12,\,12$ and $2$ neurons. In Figure~\ref{fig:annvscfd}a, a representative selection of solutions is shown. The selection comprises about $500$ solutions chosen not only from the resulting Pareto-optimal set $\pSet_\mathrm{MLP}$ but also from the last $10$ generations of NSGA-II with the best MLP used as a surrogate. Each of the solutions shown was re-evaluated by CFD.

For the continuation of the baseline framework, the common data set was used as a starting point. To achieve results comparable to those shown in Figure~\ref{fig:annvscfd}a, the NSGA-II algorithm had to be continued for $n_\mathrm{gen} = 30$ generations evaluating additional $9,600$ ejector designs. The results of the continuation are depicted in Figure~\ref{fig:annvscfd}b. Similarly to Figure~\ref{fig:annvscfd}a, the representative selection shown comprises about $500$ solutions chosen from the $\pSet_\mathrm{CFD}$ and the last $10$ generations of the continuation of NSGA-II.

Comparing the results depicted in Figures~\ref{fig:annvscfd}c and d, it may be concluded that with respect to filling the gaps in the pre-solved (common) Pareto-optimal front, CFDNNetAdapt slightly outperforms the continuation of the baseline framework. More importantly, CFDNNetAdapt was able to find geometries further expanding the common Pareto-optimal front while the baseline framework did not find similar designs, see the dashed orange box in Figure~\ref{fig:annvscfd}a. On the other hand, the new designs found by CFDNNetAdapt dominated only $55\,\%$ of the pre-solved (common) non-dominated designs. The continuation of the baseline framework was able to dominate $85\,\%$.

Regarding the computational resources used, the continuation of the baseline framework took about $6,912$ core-hours on AMD EPYC\texttrademark{}\ 7552 2nd Gen. On the other hand, CFDNNetAdapt took about $152$ core-hours to find a good MLP architecture and use it as a surrogate in an optimization. The subsequent re-evaluation of the $500$ ejector designs shown in Figure~\ref{fig:annvscfd}a took about $480$ core-hours.

\subsection{Validation of selected designs}
\label{sub:valdnn}

\begin{figure}[htbp]
    \centering
    \includegraphics[width=0.8\textwidth]{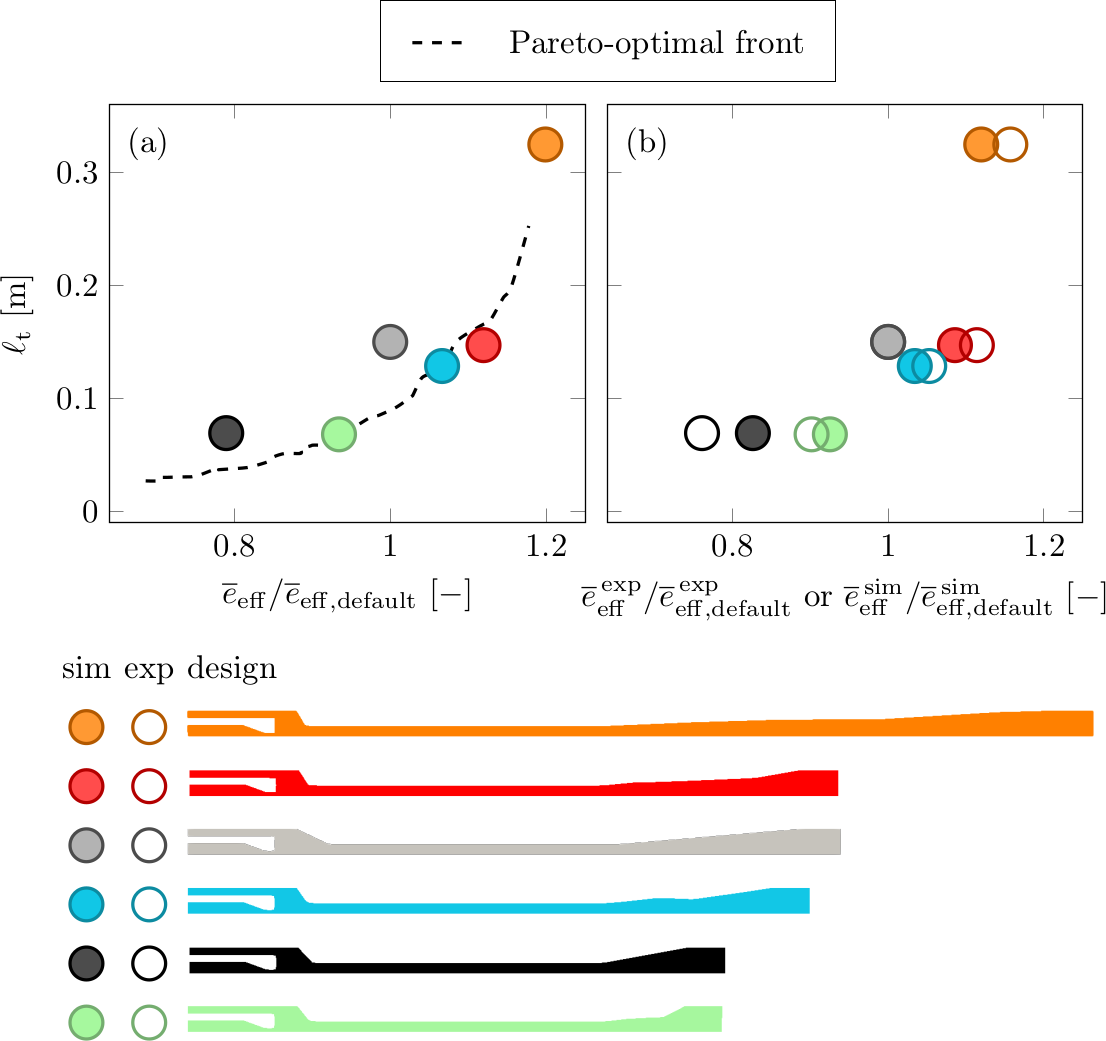}
    \caption{(a) The Pareto-optimal front depicted in Figure~\ref{fig:shapepof} with highlighted designs that were subject to experimental validation. (b) Performance of designs highlighted in (a) calculated from experimental data and corresponding CFD models.}
    \label{fig:exppof}
\end{figure}
To support the results presented in Figures~\ref{fig:shapepof} and~\ref{fig:annvscfd}, the performance of two designs found by the CFD-driven framework and two found by CFDNNetAdapt was subjected to experimental validation. The designs that were chosen are depicted in Figure~\ref{fig:exppof} in blue and green (baseline framework) and orange and red (CFDNNetAdapt). 

To complement the measurements, two more ejector designs were added to the validation plans,
\begin{inparaenum}[(i)]
    \item{the default ejector design (depicted in gray)}
    \item{and design with same component lengths as the green one but non-altered wall shapes (depicted in black).}
\end{inparaenum}
The six geometries were 3D-printed and prepared for experimental testing using the manufacturing process described in~\ref{sec:experiment}. 

With each design, measurements were made for several different primary fluid flow rates $\dot{Q}_\mathrm{inlet}$. In addition, a corresponding CFD model was constructed for each measured data point. The complete results of the validation are presented in~\ref{sec:valprint}.

The data from the measurements performed with the primary fluid flow rates closest to those used in optimization ($\dot{Q}_\mathrm{inlet} = \{0.3,\,0.4,\,0.5\}\ \mathrm{l}/\mathrm{s}$) were then used to calculate the average energy efficiency $\overline{e}_\mathrm{eff}$ by~\eqref{eq:eeff}. For each design, the efficiency obtained from optimization is given in Figure~\ref{fig:exppof}a and the efficiency calculated from experimental data ($\overline{e}_\mathrm{eff}^{\mathrm{exp}}$) and the corresponding CFD model ($\overline{e}_\mathrm{eff}^{\mathrm{sim}}$) in Figure~\ref{fig:exppof}b.

The agreement with the experimental data is not perfect, which can be attributed to the simplistic and optimization-oriented approach to CFD. Nevertheless, the trends in performance predicted by optimization hold in the experiment as well.

%%% CONCLUSIONS %%%%%%%%%%%%%%%%%%%%%%%%%%%%%%%%%%%%%%%%%%%%%%%%%%%%%%%%

\section{Summary and conclusion}
\label{sec:concl}

In this paper, we presented CFDNNetAdapt, an adaptive optimization methodology that integrates computational fluid dynamics (CFD) with multi-objective evolutionary algorithms (MOEA) and is accelerated by deep neural networks (DNNs). The methodology represents a surrogate-assisted approach to reducing computational time and enhancing the efficiency of CFD-based multi-objective optimization. CFDNNetAdapt was tested on several ZDT \revs{and LZ} benchmark functions and applied to a real-life problem of shape optimization of a single-phase ejector, the results of which were validated using experimental data.

The tests on ZDT \revs{and LZ} functions showed that CFDNNetAdapt is at the level of or outperforming similar state-of-the-art methods as SOCEMO~\citep{muller2017}, or algorithms based on Gaussian process regression~\citep{morita2022, mastrippolito2021}. Furthermore, it was listed that for sufficiently simple optimization parameter and objective space, CFDNNetAdapt should be initialized with at least $250$ data points available, but it can be expected to provide a better accuracy-to-cost ratio than NSGA-II for up to $10,000$ data points.

The application of CFDNNetAdapt to the shape optimization of an ejector further demonstrated its efficiency. This optimization represented a 14-parameter constrained problem in which the cost function was evaluated using either CFD or problem-specific penalty functions. Compared with a non-accelerated approach, CFDNNetAdapt found a wider variety of optimal ejector designs requiring a fraction of CPU time.

Future work will aim to address weak points of the approach, such as the dependency of CFDNNetAdapt on hyperparameter tuning. Additionally, the scalability of CFDNNetAdapt to other more complex real-life geometries was not fully investigated. For complex geometries, it is favorable to utilize CFD models based on an immersed boundary method for easier geometry change. However, CFDNNetAdapt in its present form is not prepared for that. \revs{Nonetheless, the current implementation of CFDNNetAdapt, which is open-source and available from \url{https://github.com/techMathGroup/CFDNNetAdapt}, was shown to significantly accelerate multi-objective optimization involving computationally expensive cost functions.}

%%% ACKNOWLEDGEMENTS %%%%%%%%%%%%%%%%%%%%%%%%%%%%%%%%%%%%%%%%%%%%%%%%%%%

\subsection*{Acknowledgments}
{
% \small
{
This research was co-funded by the European Union under the project Metamaterials for thermally stressed machine components (reg. no. CZ.02.01.01/00/23\_020/0008501). The work was financially supported by the institutional support RVO:61388998 and RVO:67985874, and the grant project with No. QL24010110 of the National Agency for Agricultural Research (NAAR). Finally, this work was supported \revs{by the Strategy AV21 programme AI: Artificial Intelligence for Science and Society, and} from the grant of Specific university research – A1\_FCHI\_2025\_004 and A1\_FCHI\_2025\_002.
}
}

%%% NOMENCLATURE %%%%%%%%%%%%%%%%%%%%%%%%%%%%%%%%%%%%%%%%%%%%%%%%%%%%%%%

\subsection*{Nomenclature}
\label{sec:nomen}
{\small
\begin{tabular}{ rl }
$A$ & plane, $[\ndUnit]$\\
$\bm{c}$ & constraints, $[\ndUnit]$\\
$d$ & diameter, $[\lUnit]$\\
$\eeff$ & energy efficiency, $[\ndUnit]$\\
$\bm{f}_\mathrm{cost}$ & cost function, $[\ndUnit]$\\
$\bm{g}$ & gravitational acceleration, $[\aUnit]$\\
$I$ & turbulence intensity, $[\ndUnit]$\\
$k$ & turbulence kinetic energy, $[\lUnit^{2}\,\tUnit^{-2}]$\\
$k_\mathrm{max}$ & maximum number of iterations, $[\ndUnit]$\\
$K_\mathrm{arm}$ & piping resistance, $[\lUnit^{-4}]$\\
$\ell$ & length, $[\lUnit]$\\
$m$ & number of objectives, $[\ndUnit]$ \\
$n_\mathrm{dnn}$ & number of deep neural networks, $[\ndUnit]$\\
$n_\mathrm{gen}$ & number of generations, $[\ndUnit]$\\
$n_\mathrm{hlrs}$ & number of hidden layers, $[\ndUnit]$\\
$n_\mathrm{pars}$ & number of parameters, $[\ndUnit]$\\
$n_\mathrm{pop}$ & population size, $[\ndUnit]$\\
$n_\mathrm{pts}$ & number of points, $[\ndUnit]$\\
$n_\mathrm{smp}$ & number of samples, $[\ndUnit]$\\
$n_\mathrm{ver}$ & number of verification cases, $[\ndUnit]$\\
$\bm{n}$ & outer unit normal, $[\ndUnit]$\\
$\bar{n}$ & average size of a hidden layer, $[\ndUnit]$\\
$\bm{o}$ & objectives, $[\ndUnit]$\\
$\pFront$ & Pareto-optimal front, $[\ndUnit]$\\
$p$ & pressure, $[\pUnit]$\\
$\tilde{p}$ & kinematic pressure, $[\kPUnit]$\\
$\bm{p}$ & parameters, $[\ndUnit]$\\
$\pSet$ & Pareto-optimal set, $[\ndUnit]$\\
$\dot{Q}$ & volumetric flow rate, $[\lUnit^{3}\,\tUnit^{-1}]$\\
$\bm{r}$ & reference point, $[\ndUnit]$\\
$r_\mathrm{area}$ & ratio of suction area in experiment and in simulation, $[\ndUnit]$\\
$r_n$ & half width of an interval of hidden layer size, $[\ndUnit]$\\
$r_Q$ & secondary-to-primary flow rate ratio, $[\ndUnit]$\\
$\mathrm{Re}$ & Reynolds number, $[\ndUnit]$\\
$s_\mathrm{cons}$ & penalty from geometry constraints, $[\ndUnit]$\\
$s_\mathrm{mesh}$ & penalty from mesh constraints, $[\ndUnit]$\\
$\bm{u}$ & velocity, $[\lUnit\,\tUnit^{-1}]$\\
$\bm{x}$ & coordinates, $[\lUnit]$\\
\end{tabular}

\subsubsection*{Greek letters}
\begin{tabular}{ rl }
$\beta$ & penalty function coefficient, {$[\ndUnit]$} \\
$\gamma$ & mesh quality indicator, {$[\ndUnit]$} \\
$\delta$ & mean error, {$[\ndUnit]$} \\
$\Delta_\mathrm{s} p $ & scaled device pressure drop, {$[\ndUnit]$} \\
$\epsilon$ & tolerance, {$[\ndUnit]$} \\
$\varepsilon$ & dissipation of turbulence kinetic energy, {$[\lUnit^{2}\,\tUnit^{-3}]$} \\
\revs{$\Lambda$} & \revs{Lebesgue measure,} {$[\ndUnit]$} \\
$\nu$ & kinematic viscosity, {$[\lUnit^{2}\,\tUnit^{-1}]$} \\
$\bm{\tau}$ & viscous stress tensor, $[\kPUnit]$\\
$\bm{\tau}'$ & Reynolds stress tensor, $[\kPUnit]$\\
$\omega$ & specific dissipation of turbulence kinetic energy, {$[\tUnit^{-1}]$} \\
$\Omega$ & computational domain, {$[\ndUnit]$} \\
\end{tabular}

\subsubsection*{Subscripts and superscripts}
\begin{tabular}{ rl}
$\mathrm{B}$ & nozzle body \\
$\mathrm{C}$ & converging part \\
$\mathrm{D}$ & diffuser \\
$\mathrm{exp}$ & experiment \\
$\mathrm{G}$ & gap part \\
$\mathrm{M}$ & mixing tube \\
$\mathrm{min}$ & minimum \\
$\mathrm{max}$ & maximum \\
$\mathrm{N}$ & nozzle \\
$\mathrm{non}$-$\mathrm{o}$ & non-orthogonality \\
$\mathrm{opt}$ & optimal \\
$\mathrm{s}$ & scaled \\
$\mathrm{sim}$ & simulation \\
$\mathrm{skew}$ & skewness \\
$\mathrm{smp}$ & samples \\
$\mathrm{t}$ & total \\
\end{tabular}

\subsubsection*{Abbreviations}
\begin{tabular}{ rl}
CFD & computational fluid dynamics \\
CFDNNetAdapt & optimization algorithm proposed in this work \\
CPU & central processing unit \\
DNN & deep neural network \\
GBM & gradient-based method \\
GPR & gaussian process regression \\
HV & hyper-volume \\
IGD & inverted generational distance \\
MLP & multi-layer perceptron \\
MOEA & multi-objective evolutionary algorithm \\
MOP & multi-objective optimization problem \\
NSGA-II & non-dominated sorting genetic algorithm II \\
RBF & radial basis function \\
SAOA & surrogate-assisted optimization algorithm \\
SOCEMO & algorithm for surrogate optimization of computationally expensive multiobjective problems \\
SM & stochastic method \\
ZDT & Zitzler-Dep-Thiele test function \\
\end{tabular}

}

\bibliography{references}

@book{OpenFOAM2007,
    author = {OpenCFD},
    title = {OpenFOAM: The Open Source CFD Toolbox. User Guide Version 1.4, OpenCFD Limited},
    year = {2007},
    publisher = {Reading UK},
}

@book{moukalled2016,
    author = {F. Moukalled and M. Darwish and L. Mangani},
    title = {The finite volume method in computational fluid dynamics: an advanced introduction with {O}pen{FOAM} and {M}atlab},
    publisher = {Springer-Verlag},
    address = {Berlin, Germany},
    edition = {1},
    year = {2016},
    isbn = {978-3-319-16874-6},
}

@article{patankar72,
    author = {S.V. Patankar and D.B. Spalding},
    title = {A calculation procedure for heat, mass and momentum transfer in three-dimensional parabolic flows},
    journal = {International Journal of Heat and Mass Transfer},
    volume = {15},
    number = {10},
    pages = {1787--1806},
    year = {1972},
    issn = {0017-9310},
    doi = {10.1016/B978-0-08-030937-8.50013-1},
}

@book{pareto1896,
    author = {V. Pareto},
    title = {Course d'Economie Politique},
    publisher = {Librairie Droz},
    place = {Geneva},
    year = {1896},
    isbn = {978-1514383506},
}

@book{goldbergBook1989,
    author = {D.E. Goldberg},
    title = {Genetic Algorithms in Search, Optimization, and Machine Learning},
    place = {Alabama},
    publisher = {Addison-Wesley publishing company, inc.},
    year = {1989},
    isbn = {978-0201157673},
}

@article{emmerich2018,
    author = {M.T.M. Emmerich and A.H. Deutz},
    title = {A tutorial on {M}ultiobjective {O}ptimization: {F}undamentals and {E}volutionary {M}ethods},
    journal = {Natural Computing},
    volume = {17},
    pages = {585--609},
    year = {2018},
    doi = {10.1007/s11047-018-9685-y},
}

@book{deb2001,
    author = {K. Deb},
    title = {Multi-objective Optimization Using Evolutionary Algorithms},
    publisher = {Wiley},
    place = {Chichester},
    year = {2001},
    isbn = {978-0470743614},
}

@article{zitzler2000,
    author = {E. Zitzler and K. Deb and L. Thiele},
    journal = {Evolutionary Computation},
    title = {Comparison of Multiobjective Evolutionary Algorithms: {E}mpirical Results},
    year = {2000},
    volume = {8},
    number = {2},
    pages = {173--195},
    doi = {10.1162/106365600568202},
}

@article{deb2002,
    author = {K. Deb and A. Pratap and S. Agarwal and T. Meyarivan},
    title = {A fast and elitist multiobjective genetic algorithm: {NSGA}-{II}},
    journal = {I{EEE} Transactions on Evolutionary Computation},
    year = {2002},
    volume = {6},
    number = {2},
    pages = {182--197},
    doi = {10.1109/4235.996017},
}

@article{daniels2019,
    author = {S.J. Daniels and A.A.M. Rahat and G.R. Tabor and J.E. Fieldsend and R.M. Everson},
    title = {Automated shape optimization of a plane asymmetric diffuser using combined {C}omputational {F}luid {D}ynamic simulations and multi-objective {B}ayesian methodology},
    journal = {International Journal of Computational Fluid Dynamics},
    volume = {33},
    pages = {256--271},
    year = {2019},
    doi = {10.1080/10618562.2019.1683165},
}

@article{havelka1997,
    author = {P. Havelka and V. Linek and J. Sinkule and J. Zahradnik and M. Fialova},
    title = {Effect of the ejector configuration on the gas suction rate and gas hold-up in ejector loop reactors},
    journal = {Chemical Engineering Science},
    volume = {52},
    pages = {1701--1713},
    year = {1997},
    doi = {10.1016/S0009-2509(97)00003-1}
}

@article{taleghami2018,
    author = {S.T. Taleghani and M. Sorin and S. Poncet},
    title = {Modeling of two-phase transcritical {CO}$_2$ ejectors for on-design and off-design conditions},
    journal = {International Journal of Refrigeration},
    volume = {87},
    pages = {91--105},
    year = {2018},
    doi = {10.1016/j.ijrefrig.2017.10.025},
}

@software{platypus,
    author = {{D. Hadka}},
    title = {Platypus, {A} {F}ree and {O}pen {S}ource {P}ython {L}ibrary for {M}ultiobjective {O}ptimization},
    url = {https://github.com/Project-Platypus/Platypus},
    version = {1.0.4},
    date = {2020-04-14},
    year = {2020},
}

@software{pyrenn,
    author = {{D. Atabay, Institute for Energy Economy and Application Technology, Technische Universität München}},
    title = {{pyrenn: A recurrent neural network toolbox for python and matlab}},
    url = {https://pyrenn.readthedocs.io/en/latest/},
    version = {0.1},
    date = {2018-06-30},
    year = {2018},
}

@article{gpflow2017,
    author = {A.G. de G. Matthews and M. van der Wilk and T. Nickson and K. Fujii and A. Boukouvalas and P. Le{\'o}n-Villagr{\'a} and Z. Ghahramani and J. Hensman},
    title = {{{GP}flow: A {G}aussian process library using {T}ensor{F}low}},
    journal = {Journal of Machine Learning Research},
    month = {04},
    volume = {18},
    number = {40},
    pages = {1--6},
    year = {2017},
    doi = {10.48550/arXiv.1610.08733},
}

@TechReport{menter1992,
    author = {F.R. Menter},
    title = {Improved two equation k-$\omega$ turbulence models for aerodynamic flows},
    number = {N93-22809},
    institution = {NASA},
    year = {1992},
    pages = {34},
}

@InProceedings{hellsten1997,
    author = {A. Hellsten},
    title = {Some improvements in {M}enter's k-$\omega$ {SST} turbulence model},
    booktitle = {Proceedings of the Fluid Dynamics Conference},
    publisher = {AIAA},
    location = {Albuquerque, USA},
    year = {1997},
    volume = {71},
    pages = {1--11},
    doi = {10.2514/6.1998-2554},
}

@article{launder1974,
    author = {B.E. Launder and D.B. Spalding},
    title = {The numerical computation of turbulent flows},
    journal = {Computer Methods in Applied Mechanics and Engineering},
    volume = {3},
    pages = {269--289},
    doi = {10.1016/0045-7825(74)90029-2},
    year = {1974},
}

@article{elTahry1983,
    author = {S.H. El Tahry},
    title = {K-$\epsilon$ equation for compressible reciprocating engine flows},
    journal = {Journal of Energy},
    volume = {7},
    pages = {345--353},
    year = {1983},
    doi = {10.2514/3.48086}
}

@article{shih1995,
    author = {T.H. Shih and W.W. Liou and A. Shabbir and Z. Yang and J. Zhu},
    title = {A new k-$\epsilon$ eddy viscosity model for high {R}eynolds number turbulent flows},
    journal = {Computers \& Fluids},
    volume = {24},
    pages = {227--238},
    year = {1995},
    doi = {10.1016/0045-7930(94)00032-T},
}

@book{wilcox2006,
    author = {D.C. Wilcox},
    title = {Turbulence modeling for CFD},
    year = {2006},
    edition = {3},
    publisher = {DCW Industries, USA},
    isbn = {978-1928729082},
}

@article{russo2016,
    author = {F. Russo and N.T. Basse},
    title = {Scaling of turbulence intensity for low-speed flow in smooth pipes},
    journal = {Flow Measurement and Instrumentation},
    year = {2016},
    volume = {52},
    pages = {101--114},
    doi = {10.1016/j.flowmeasinst.2016.09.012},
}

@book{goodfellow2016,
    author = {I. Goodfellow and Y. Bengio and A. Courville},
    title = {Deep Learning},
    publisher = {MIT Press},
    year = {2016},
    isbn = {978-0262035613},
}

@article{li2017,
    author = {Z. Li and X. Zheng},
    title = {Review of design optimization methods for turbomachinery aerodynamics},
    journal = {Progress in Aerospace Sciences},
    year = {2017},
    volume = {93},
    pages = {1--23},
    doi = {10.1016/j.paerosci.2017.05.003},
}

@article{skinner2018,
    author = {S.N. Skinner and H. Zare-Behtash},
    title = {State-of-the-art in aerodynamic shape optimisation methods},
    journal = {Applied Soft Computing},
    year = {2018},
    volume = {62},
    pages = {933--962},
    doi = {10.1016/j.asoc.2017.09.030},
}

@article{foster1997,
    author = {N.F. Foster and G.S. Dulikravich},
    title = {Three-dimensional aerodynamic shape optimization using genetic and gradient search algorithms},
    journal = {Journal of Spacecraft and Rockets},
    year = {1997},
    volume = {34},
    pages = {36--42},
    doi = {10.2514/2.3189},
}

@article{gill2005,
    author = {P.E. Gill and W. Murray and M.A. Saunders},
    title = {{SNOPT:} An {SQP} Algorithm for large-scale constrained optimization},
    journal = {SIAM Review},
    year = {2005},
    volume = {47},
    pages = {99--131},
    doi = {10.1137/S0036144504446096},
}

@article{svanberg1987,
    author = {K. Svanberg},
    title = {The method of moving asymptotes -- A new method for structural optimization},
    journal = {International Journal for Numerical Methods in Engineering},
    year = {1987},
    volume = {24},
    pages = {359--373},
    doi = {10.1002/nme.1620240207},
}

@inproceedings{eberhart1995,
    author = {R. Eberhart and J. Kennedy},
    booktitle = {MHS'95. Proceedings of the Sixth International Symposium on Micro Machine and Human Science}, 
    title = {A new optimizer using particle swarm theory}, 
    year = {1995},
    volume = {},
    number = {},
    pages = {39--43},
    doi = {10.1109/MHS.1995.494215}
}

@article{kirkpatrick1983,
    author = {S. Kirkpatrick  and C.D. Gelatt  and M.P. Vecchi },
    title = {Optimization by Simulated Annealing},
    journal = {Science},
    volume = {220},
    number = {4598},
    pages = {671--680},
    year = {1983},
    doi = {10.1126/science.220.4598.671},
}

@article{yu2018,
    author = {Y. Yu and Z. Lyu and Z. Xu and J.R.R.A. Martins},
    title = {On the influence of optimization algorithm and initial design on wing aerodynamic shape optimization},
    journal = {Aerospace Science and Technology},
    volume = {75},
    pages = {183--199},
    year = {2018},
    doi = {10.1016/j.ast.2018.01.016},
}

@article{kim2014,
    author = {J.-H. Kim and B. Ovgor and K.-H. Cha and J.-H. Kim and S. Lee and K.-Y. Kim},
    title = {Optimization of the aerodynamic and aeroacoustic performance of an axial-flow fan},
    journal = {AIAA Journal},
    volume = {52},
    pages = {2032--2044},
    year = {2014},
    doi = {10.2514/1.J052754},
}

@article{vicini1999,
    author = {A. Vicini and D. Quagliarella},
    title = {Airfoil and wing design through hybrid optimization strategies},
    journal = {AIAA Journal},
    volume = {37},
    pages = {634--641},
    year = {1999},
    doi = {10.2514/2.764},
}

@article{gage1995,
    author = {P.J. Gage and I.M. Kroo and I.P. Sobieski},
    title = {Variable-complexity genetic algorithm for topological design},
    journal = {AIAA Journal},
    volume = {33},
    pages = {2212--2217},
    year = {1995},
    doi = {10.2514/3.12969},
}

@article{roe1986,
    author = {P.L. Roe},
    title = {Characteristics-based schemes for the {Euler} equations},
    year = {1986},
    volume = {18},
    journal = {Annual Reviews of Fluid Mechanics},
    pages = {337--365},
    doi = {10.1146/annurev.fl.18.010186.002005},
}

@inproceedings{kubickova2021,
    author = {L. Kubíčková and M. Isoz and J. Haidl},
    title = {Increasing Ejector Efficiency via Diffuser Shape Optimization},
    booktitle = {Proceedings of Topical Problems of Fluid Mechanics 2021},
    editor = {D. Šimurda and T. Bodnár},
    publisher = {IT CAS},
    year = {2021},
    month = {02},
    pages = {79--86},
    doi = {10.14311/TPFM.2021.011}
}

@article{levenberg1944,
    author = {K. Levenberg},
    title = {A Method for the Solution of Certain Problems in Least Squares},
    volume = {2},
    journal = {Quarterly of Applied Mathematics},
    pages = {164--168},
    year = {1944},
    doi = {10.1090/qam/10666}
}

@article{marquardt1963,
    author = {D. Marquardt},
    title = {An Algorithm for Least-Squares Estimation of Nonlinear Parameters},
    volume = {11},
    journal = {Journal of the Society for Industrial and Applied Mathematics},
    pages = {431--441},
    year = {1963},
    doi = {10.1137/0111030},
}

@article{gebousky2023,
    author = {O. Gebouský and K. Mařík and J. Haidl and M. Zedniková},
    title = {Enhancement of gas entrainment rate in liquid-gas ejector pump},
    journal = {Chemical Engineering Research and Design},
    volume = {189},
    pages = {117--125},
    year = {2023},
    doi = {10.1016/j.cherd.2022.11.009},
}

@article{utomo2009,
    author = {T. Utomo and Z. Jin and M. Rahman and H. Jeong and H. Chung},
    title = {Investigation on hydrodynamics and mass transfer characteristics of a gas-liquid ejector using three-dimensional {CFD} modeling},
    journal = {Journal of Mechanical Science and Technology},
    month = {09},
    pages = {1821--1829},
    volume = {22},
    year = {2009},
    doi = {10.1007/s12206-008-0614-3},
}

@software{oF,
    author = {H. Weller and C. Greenshields and W. Bainbridge et al.},
    title = {Open{FOAM}},
    url = {https://openfoam.org/},
    version = {8},
    date = {2022-07-22},
}

@article{morita2022,
    title = {Applying Bayesian optimization with Gaussian process regression to computational fluid dynamics problems},
    author = {Y. Morita and S. Rezaeiravesh and N. Tabatabaei and R. Vinuesa and K. Fukagata and P. Schlatter},
    journal = {Journal of Computational Physics},
    volume = {449},
    pages = {110788},
    year = {2022},
    issn = {0021-9991},
    doi = {10.1016/j.jcp.2021.110788},
}

@article{mastrippolito2021,
    title = {Kriging metamodels-based multi-objective shape optimization applied to a multi-scale heat exchanger},
    author = {F. Mastrippolito and S. Aubert and F. Ducros},
    journal = {Computers \& Fluids},
    volume = {221},
    pages = {104899},
    year = {2021},
    issn = {0045-7930},
    doi = {10.1016/j.compfluid.2021.104899},
}

@article{guerreiro2021,
    title = {The Hypervolume Indicator: {C}omputational Problems and Algorithms},
    author = {A.P. Guerreiro and C.M. Fonseca and L. Paquete},
    journal = {{A}{C}{M} computing surveys},
    volume = {54},
    pages = {42},
    year = {2021},
    doi = {10.1145/3453474},
}

@article{dong2021,
    author = {Huachao Dong and Jinglu Li and Peng Wang and Baowei Song and Xinkai Yu},
    title = {Surrogate-guided multi-objective optimization ({SGMOO}) using an efficient online sampling strategy},
    journal = {Knowledge-Based Systems},
    volume = {220},
    pages = {106919},
    year = {2021},
    issn = {0950-7051},
    doi = {10.1016/j.knosys.2021.106919},
}

@article{muller2017,
    title={{SOCEMO}: {S}urrogate Optimization of Computationally Expensive Multiobjective Problems},
    author={J. M{\"u}ller},
    journal={INFORMS J. Comput.},
    year={2017},
    volume={29},
    pages={581--596},
    doi={10.1287/ijoc.2017.0749},
}

@article{manriquez2016,
    author = {A. Díaz-Manríquez and G. Toscano and J.H. Barron-Zambrano and E. Tello-Leal},
    title = {A Review of Surrogate Assisted Multiobjective Evolutionary Algorithms},
    journal = {Computational Intelligence and Neuroscience},
    volume = {2016},
    number = {1},
    pages = {9420460},
    doi = {10.1155/2016/9420460},
    year = {2016}
}

@inproceedings{uhrenholt2021,
    author = {A.K. Uhrenholt and V. Charvet and B.S. Jensen},
    title = {Probabilistic Selection of Inducing Points in Sparse Gaussian Processes},
    booktitle = {Proceedings of the Thirty-Seventh Conference on Uncertainty in Artificial Intelligence (UAI 2021)},
    month = {},
    pages = {1035--1044},
    year = {2021},
    doi = {10.48550/arXiv.2010.09370}
}

@article{chugh2016,
  author = {T. Chugh and Y. Jin and K. Miettinen and J. Hakanen and K. Sindhya},
  title = {A Surrogate-Assisted Reference Vector Guided Evolutionary Algorithm for Computationally Expensive Many-Objective Optimization}, 
  journal = {{IEEE} Transactions on Evolutionary Computation}, 
  year = {2018},
  volume = {22},
  number = {1},
  pages = {129--142},
  doi = {10.1109/TEVC.2016.2622301}
}

@article{knowles2006,
  author = {J. Knowles},
  title = {Par{EGO}: {A} hybrid algorithm with on-line landscape approximation for expensive multiobjective optimization problems}, 
  journal = {{IEEE} Transactions on Evolutionary Computation}, 
  year = {2006},
  volume = {10},
  number = {1},
  pages = {50--66},
  doi = {10.1109/TEVC.2005.851274}
}

@article{tian2023,
    author = {Y. Tian and J. Hu and C. He and H. Ma and L. Zhang and X. Zhang},
    title = {A pairwise comparison based surrogate-assisted evolutionary algorithm for expensive multi-objective optimization},
    journal = {Swarm and Evolutionary Computation},
    volume = {80},
    pages = {101323},
    year = {2023},
    doi = {10.1016/j.swevo.2023.101323},
}

@article{wolday2024,
    author = {A.K. Wolday and M. Ramteke},
    title = {Surrogate model-based optimization of methanol synthesis process for multiple objectives: {A} pathway towards achieving sustainable development goals},
    journal = {Chemical Engineering Research and Design},
    volume = {204},
    pages = {172--182},
    year = {2024},
    doi = {10.1016/j.cherd.2024.02.021},
}

@article{marini1992,
    author = {M. Marini and A. Massardo and A. Satta and M. Geraci},
    title = {Low Area Ratio Aircraft Fuel Jet Pump Performances With and Without Cavitation},
    journal = {Journal of Fluids Engineering},
    volume = {114},
    number = {4},
    pages = {626--631},
    year = {1992},
    doi = {10.1115/1.2910077},
}

@article{jones1998,
    author = {D.R. Jones and M. Schonlau and W.J. Welch},
    title = {Efficient Global Optimization of Expensive Black-Box Functions},
    journal = {Journal of Global Optimization},
    volume = {13},
    number = {},
    pages = {455--492},
    year = {1998},
    doi = {10.1023/A:1008306431147},
}

@article{li2009,
    author = {H. Li and Q. Zhang},
    title = {Multiobjective Optimization Problems With Complicated Pareto Sets, {MOEA}/{D} and {NSGA}-{II}},
    journal = {{IEEE} Transactions on Evolutionary Computation},
    volume = {13},
    number = {2},
    pages = {284--302},
    year = {2009},
    doi = {10.1109/TEVC.2008.925798},
}

@inproceedings{fonseca2006,
    author = {C.M. Fonseca and L. Paquete and M. Lopez-Ibanez},
    title = {An Improved Dimension-Sweep Algorithm for the Hypervolume Indicator},
    booktitle = {Proceedings of the {IEEE} International Conference on Evolutionary Computation},
    month = {7},
    pages = {3973--3979},
    year = {2006},
    doi = {10.1109/CEC.2006.1688440}
}

@article{zhu2024,
    author = {H. Zhu and L. Shi and Z. Hu and Q. Su},
    title = {A multi-surrogate multi-tasking genetic algorithm with an adaptive training sample selection strategy for expensive optimization problems},
    journal = {Engineering Applications of Artificial Intelligence},
    volume = {130},
    pages = {107684},
    year = {2024},
    doi = {10.1016/j.engappai.2023.107684},
}

@article{deb2014,
  author = {K. Deb and H. Jain},
  title = {An Evolutionary Many-Objective Optimization Algorithm Using Reference-Point-Based Nondominated Sorting Approach, Part I: Solving Problems With Box Constraints}, 
  journal = {IEEE Transactions on Evolutionary Computation}, 
  year = {2014},
  volume = {18},
  number = {4},
  pages = {577--601},
  doi = {10.1109/TEVC.2013.2281535}
}

\begin{appendix}
    \clearpage
    \setcounter{figure}{0}
    \setcounter{table}{0}
    \section{Experimental set-up and default geometry}
    \label{sec:experiment}
    Measurements were performed on a modular laboratory-scale ejector unit, allowing liquid-liquid and liquid-gas process variants to be used. In the present study, water at ambient temperature was used as both the primary fluid (driving) and the secondary fluid (entrained), that is, the ejector was operated in the liquid-liquid regime.
    
    \begin{figure}[htbp]
        \centering
        \includegraphics{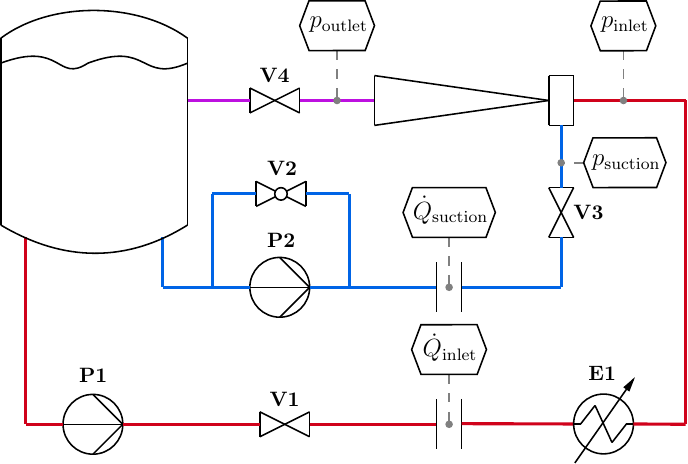}
        \caption{Sketch of the experimental setup. P1,P2 - centrifugal pumps; E1 - heat exchanger; V1-V4 - valves; red lines - primary (driving) fluid; blue lines - secondary (entrained) fluid; magenta lines - mixed stream.}
        \label{fig:expSetup}
    \end{figure}
    
    \begin{figure}[htbp]
        \centering
        \includegraphics{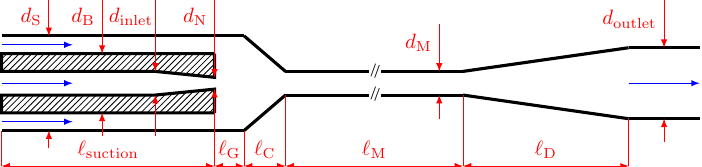}
        \caption{Annotated drawing of the default ejector geometry.}
        \label{fig:ejectorDrawing}
    \end{figure}
    
    \begin{table}[htbp]
        \centering
        \begin{tabular}{cc|cc|cc}
            dim. & value [mm] & dim. & value [mm] & dim. & value [mm] \\
            \midrule
            $d_\mathrm{suction}$ & $35$ & $d_\mathrm{B}$ & $25$ & $d_\mathrm{inlet}$ & $16$\\
            $d_\mathrm{N}$ & $\{4.3,\,5.0,\,7.0\}$ & $d_\mathrm{M}$ & $\{14,\,19\}$ & $d_\mathrm{outlet}$ & $35.6$\\
            $\ell_\mathrm{suction}$ & $60$ & $\ell_\mathrm{G}$ & $15$ & $\lenC$ & $20$\\
            $\ell_\mathrm{M}$ & $200$ & $\lenD$ & $130$ & $-$ & $-$\\
        \end{tabular}
        \caption{Default values of ejector dimensions depicted in Figure~\ref{fig:ejectorDrawing}.}
        \label{tab:ejectorDims}
    \end{table}
    
    The experimental setup is sketched in Figure~\ref{fig:expSetup}. The setup comprises a 60~L storage tank from which water is pumped using the centrifugal pump P1 through the control valve V1, a heat exchanger E1, and a nozzle to the suction chamber of the ejector unit sketched in Figure~\ref{fig:ejectorDrawing} with dimensions given in Table~\ref{tab:ejectorDims}. To allow independent pressure control in the suction chamber $p_\mathrm{suction}$, the secondary fluid is delivered using a second centrifugal pump P2; the pressure in the suction chamber is adjusted by the throttling valve V3. Similarly, the pressure difference between the suction chamber and the diffuser outlet is controlled by the throttling valve V4.
    
    Primary ($\dot{Q}_\mathrm{inlet}$) and secondary fluid flow rates ($\dot{Q}_\mathrm{suction}$) were measured as the pressure difference of the calibrated orifices; the relative uncertainty of the flow rate is less than 1\% of the reading. The pressure in front of the nozzle ($p_\mathrm{inlet}$) was measured using the Cressto SR transducer calibrated for the relative pressure range from -100~kPa to $1,000$~kPa with the uncertainty of $\pm$5~kPa; for the measurement of the pressures in the suction chamber ($p_\mathrm{suction}$) and at the diffuser outlet ($p_\mathrm{outlet}$), the Cressto SR transducers were calibrated for the relative pressure range from -100~kPa to 300~kPa with the uncertainty of $\pm$0.75~kPa. 
    
    To allow for affordable CFD simulations while providing relevant validation data, the ejector unit was designed to be axially symmetric, including the suction chamber. Furthermore, the modular character of the ejector unit enables the alteration of the unit geometry by changing its individual parts - nozzle, converging part, mixing tube, and diffuser. In this work, data were measured using several unit geometries. The one sketched in Figure~\ref{fig:ejectorDrawing} with parts dimensions from Table~\ref{tab:ejectorDims} was the original configuration of the liquid-gas ejector unit used in~\citep{gebousky2023} and it is referred to as the default geometry. Other used configurations resulted from the geometry optimization of the shape of the converging part and the diffuser. In optimization, only geometries with $d_\mathrm{N} = 4.3$~mm and $d_\mathrm{M} = 14$~mm were considered. The tested 'optimal' configurations, differing substantially in the total length of the ejector unit, were selected from the Pareto-optimal set.
    
    In all cases, the converging part and diffuser were manufactured from polylactic acid (PLA) using an FDM 3D printer, 0.4~mm nozzle, and 0.2~mm layer height. The inner surface of the printed parts was polished with 1200 grit sandpaper and impregnated using an acrylic varnish. The manufacturing process produced airtight parts with surface smoothness comparable to that of stainless steel piping.
    
    \setcounter{figure}{0}
    \setcounter{table}{0}
    \section{Mathematical model}
    \label{sec:mathMod}
    A mathematical model of a single phase (water-water) ejector was prepared in the open-source C++ CFD library OpenFOAM (OpenFOAM Foundation version 8)~\citep{OpenFOAM2007}. The geometry of the physical device and the experimental setup are described in \ref{sec:experiment}. The corresponding CFD model was constructed with an emphasis on being computationally affordable and fully automated, including geometry generation and meshing. In the following, we present the flow governing equations considered, the applied boundary conditions and initial guess.
    
    \subsection{Governing equations}
    \label{sec:goveq}
    The flow in the water-water ejector was simulated to be pseudo-stationary, axisymmetric, isothermic and incompressible. Furthermore, water was assumed to be a Newtonian fluid, and the effects of turbulence were taken into account by Reynolds averaging. For the given case, the Reynolds-averaged Navier-Stokes (RANS) equations take the form of
    \begin{equation}
    \label{eq:RANS}
        \begin{array}{*1{>{\displaystyle}r}*1{>{\displaystyle}c}*1{>{\displaystyle}l}}
            \nabla\cdot \left( \overline{\bm{u}}\otimes \overline{\bm{u}} \right) - \nabla \cdot \left(\overline{\bm{\tau}} + \bm{\tau}' \right)  &=&  -\nabla \overline{\tilde{p}} + \bm{g}\\[0.2cm]
            \nabla \cdot \overline{\bm{u}} &=& 0\,,
        \end{array}
    \end{equation}
    where $\overline{\bm{u}}$ and $\overline{\tilde{p}}$ are the averaged velocity and kinematic pressure, respectively. Next, $\overline{\bm{\tau}} = \nu \nabla \overline{\bm{u}}$ is the averaged viscous stress tensor with $\nu$ being the fluid kinematic viscosity. The second stress tensor, $\bm{\tau}' = \bm{u}' \otimes \bm{u}'$, corresponds to the so-called Reynolds stress tensor where $\bm{u}'$ is the instantaneous turbulence-driven velocity fluctuation. Finally, $\bm{g}$ is the gravitational acceleration. Note that, hereafter, only the Reynolds-averaged variables will be taken into account and the bar over individual symbols will be omitted.
    
    Due to the presence of the Reynolds stress term $(\nabla \cdot \bm{\tau}')$, the formulation~\eqref{eq:RANS} requires an additional closure model. Four two-equation eddy viscosity closure models were tested and chosen from, namely the $k$-$\varepsilon$ model of~\citet{launder1974} with rapid distortion theory (RDT) based compression term~\citep{elTahry1983}, realizable $k$-$\varepsilon$ model by~\citet{shih1995}, $k$-$\omega$ model of~\citet{wilcox2006} and the Menter model $k$-$\omega$ shear stress transport (SST) model~\citep{menter1992} in the formulation given by Hellsten~\citep{hellsten1997}.
    
    \subsection{Boundary conditions and initial guess}
    \label{sec:bcic}
    Both the system~\eqref{eq:RANS} and the selected turbulence closure model need to be completed with suitable boundary conditions. We divide the computational domain boundary into
    \begin{inparaenum}[(i)]
        \item{\textit{inlet} boundary representing the primary fluid inlet,}
        \item{\textit{suction} boundary that corresponds to the secondary fluid inlet via the suction chamber,}
        \item{\textit{outlet} boundary as the outlet from the ejector and}
        \item{\textit{wall} boundary representing the system solid boundaries.}
    \end{inparaenum}
    The types of boundary conditions used for each section are given in Table~\ref{tab:bcs}.
    
    \begin{table}[htbp]
    \centering
    {
    \begin{tabular}{ccc}
    \toprule
        \multicolumn{3}{c}{\textit{inlet}} \\
    \midrule
        $\bm{u}$ & $\tilde{p}$ & $k\,|\,\omega\,|\,\varepsilon$ \\
        fixed value & zero gradient & fixed value \\
    \midrule
        \multicolumn{3}{c}{\textit{outlet}} \\
    \midrule
        $\bm{u}$ & $\tilde{p}$ & $k\,|\,\omega\,|\,\varepsilon$ \\
        inlet-outlet & total pressure & inlet-outlet \\
    \midrule
        \multicolumn{3}{c}{\textit{suction}} \\
    \midrule
        $\bm{u}$ & $\tilde{p}$ & $k\,|\,\omega\,|\,\varepsilon$ \\
        inlet-outlet & total pressure with resistance & inlet-outlet \\
    \midrule
        \multicolumn{3}{c}{\textit{wall}} \\
    \midrule
        $\bm{u}$ & $\tilde{p}$ & $k\,|\,\omega\,|\,\varepsilon$ \\
        no-slip & zero gradient & wall functions \\
    \bottomrule
    \end{tabular}
    }
    \caption{Applied boundary conditions.}
    \label{tab:bcs}
    \end{table}
    
    Most of the boundary conditions used are implemented by default in the OpenFOAM library~\citep{OpenFOAM2007}. The only exception is the boundary condition for the kinematic pressure at the suction boundary. An additional resistance term was added to reflect the hydraulic behavior of the experimental setup. In particular, the resistance is dependent on the secondary fluid flow rate ($\dot{Q}_{\mathrm{suction}}$) and the final form of the boundary condition is
    \begin{equation}
    \label{eq:BCSuctionPressure}
        \tilde{p} = \tilde{p}^0 -\frac{1}{2}\norm{\bm{u}}^{2} + K_{\mathrm{arm}}\, r_\mathrm{area}\, \dot{Q}_{\mathrm{suction}}^{2}\,,
    \end{equation}
    with $\tilde{p}^0$ being the total pressure, $K_{\mathrm{arm}}$ a piping resistance determined from the experimental data and $r_\mathrm{area}$ a ratio of the suction area available in experiment to the area of the \textit{suction} boundary.
    
    Finally, the problem specification was completed by prescribing a standard initial guess based on formulas by~\citet{russo2016}.
    
    \subsection{Numerical settings}
    \label{sub:numerics}
    The flow governing equations~\eqref{eq:RANS} and the turbulence closure models tested were discretized within the C++ finite-volume library OpenFOAM~\citep{OpenFOAM2007}. The overall solution method was a segregated steady-state one. In particular, the pressure and velocity fields were solved by using the SIMPLE loop~\citep{patankar72} with modifications described in~\citet{moukalled2016}, while the turbulence variables were updated at the end of each solver iteration. An emphasis was made on keeping the problem discretization as close to the second-order accuracy as possible. To do so, a second-order upwind scheme was used for discretization of the momentum equation convection term. For the convection of turbulence variables, a high-resolution scheme with the Minmod limiter by~\citet{roe1986} was utilized. For automatic simulation termination, the convergence criteria were set to $10^{-4}$ for all fields.
    
    \subsection{Mesh generation and mesh size independence study}
    \label{sub:meshSize}
    
    \begin{figure}[htbp]
        \centering
        \includegraphics{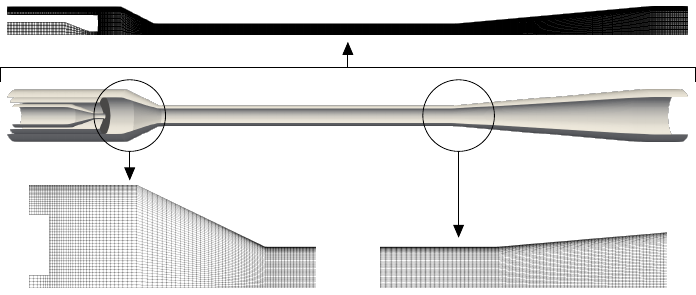}
        \caption{Ejector geometry and qualitative view of the computational mesh. The overall mesh structure is shown in the top part of the figure, details of specific positions along the ejector are shown on the figure bottom.}
        \label{fig:mesh}
    \end{figure}
    
    The computational domain was based on a 2D axi-symmetric approximation of the ejector device. The computational mesh used is depicted in Figure~\ref{fig:mesh}. The mesh was refined close to the walls, in order to keep the $y^+$ measure below $1$. With the default ejector geometry (see Table~\ref{tab:ejectorDims}), the maximum non-orthogonality of the mesh was 25 degrees and the maximum skewness of the mesh was 0.6.
    
    \begin{figure}[htbp]
        \centering
        \includegraphics{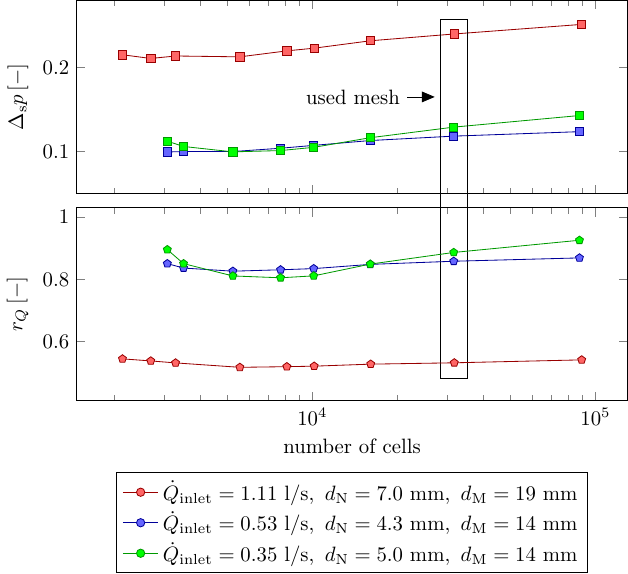}
        \caption{Dependence of the scaled device pressure drop and secondary-to-primary flow rate ratio on the mesh resolution. Simulations were computed using the $k$-$\omega$ SST turbulence model.}
        \label{fig:meshSize}
    \end{figure}
    
    To investigate the dependence of the results on the mesh resolution, two indicators were chosen. The first one was a scaled pressure drop of the device that was defined as
    \begin{equation}
        \label{eq:scaledDPApp}
        \Delta_\mathrm{s} p := \frac{\overline{p}_{\mathrm{outlet}} - \overline{p}_{\mathrm{suction}}}{\overline{p}_{\mathrm{inlet}} - \overline{p}_{\mathrm{outlet}}}\,,
    \end{equation}
    where $p$ stands for the kinematic pressure and the bars over variables denote surface-averages. The second indicator was a secondary-to-primary flow rate ratio computed as
    \begin{equation}
        \label{eq:flowratio}
        r_Q = \dot{Q}_\mathrm{suction}/\dot{Q}_\mathrm{inlet}
    \end{equation}
    where $\dot{Q}$ is the fluid flow rate. The same indicators were later used for model validation against experimental data. Note that the mesh size independence study was performed using the $k$-$\omega$ SST turbulence model.
    
    Results of the mesh size independence study are shown in Figure~\ref{fig:meshSize}. Each data series corresponds to an ejector geometry with specified outer diameter of the nozzle ($d_\mathrm{N}$) and diameter of the mixing tube ($d_\mathrm{M}$). Each geometry was operated with different primary fluid flow rates ($\dot{Q}_\mathrm{inlet}$) to cover the entire range of experimental data.
    
    With increasing mesh resolution, there is a weak trend of increasing $\Delta_\mathrm{s} p$ and $r_Q$. However, the indicators mostly level-out for meshes with more than $10^4$ cells. Hence, the mesh having $\sim 3\times10^4$ cells (framed results in Figure~\ref{fig:meshSize}) was selected as the one providing the most suitable trade-off between the solution accuracy and computational cost.
    
    \begin{figure}[htbp]
        \centering
        \includegraphics[width=0.8\textwidth]{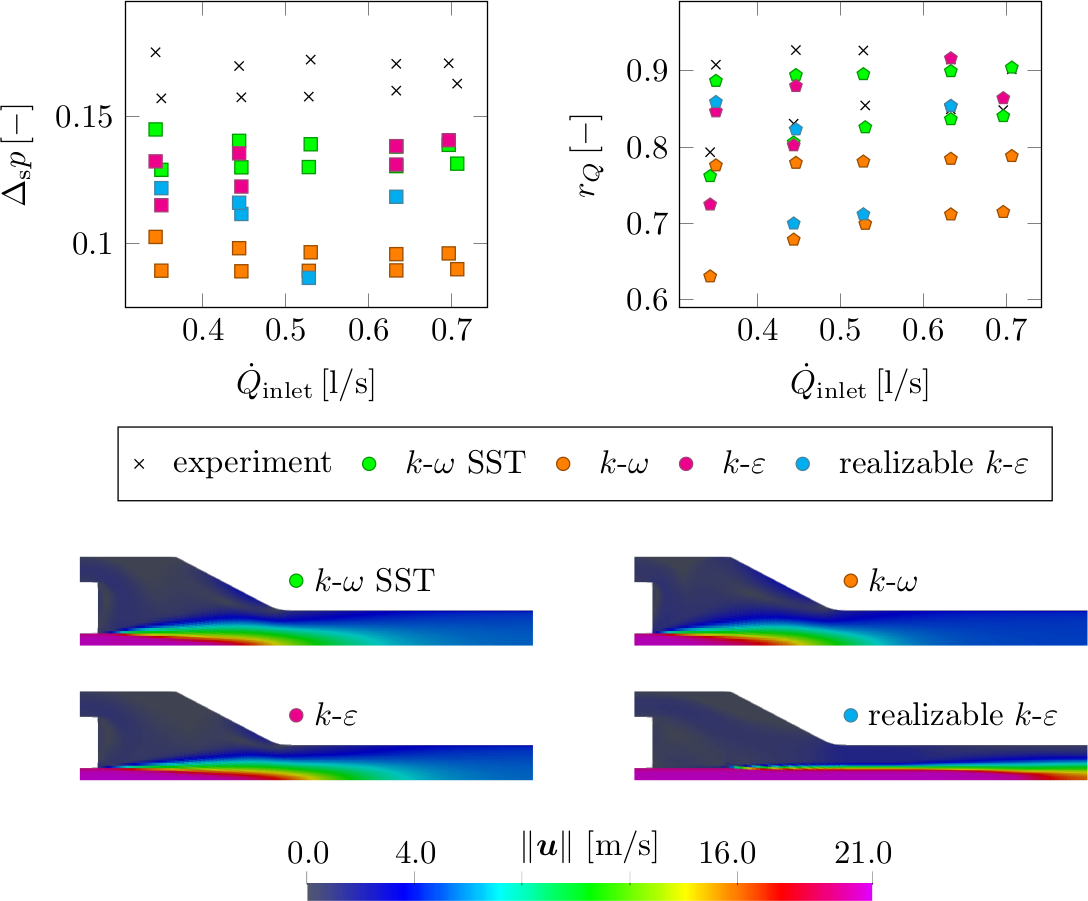}
        \caption{Comparison of experimental and simulation data computed with different turbulence models for the ejector geometry with $d_\mathrm{N} = 5.0\ \mathrm{mm}$ and $d_\mathrm{M} = 14\ \mathrm{mm}$. Velocity fields predicted by each turbulence model are given at the figure bottom.}
        \label{fig:turbms}
    \end{figure}
    
    \subsection{Turbulence model selection}
    \label{sub:turbModel}
    Experimental data measured on a geometry with $d_\mathrm{N} = 5.0\ \mathrm{mm}$ and $d_\mathrm{M} = 14\ \mathrm{mm}$ were used to test different turbulence models, namely the $k$-$\omega$ SST, $k$-$\omega$, $k$-$\varepsilon$ and realizable $k$-$\varepsilon$ model. The results are presented in Figure~\ref{fig:turbms}.
    
\begin{figure}[htbp]
        \centering
        \includegraphics{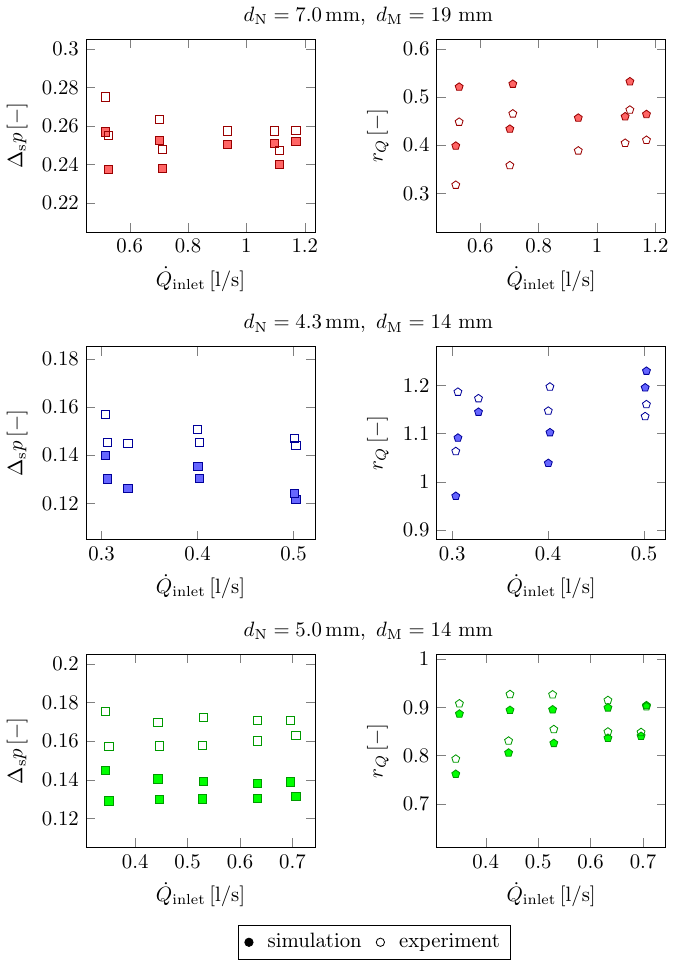}
        \caption{Comparison of simulation and experimental data, namely the scaled device pressure drop ($\Delta_\mathrm{s} p$) and secondary-to-primary flow rate ratio ($r_Q$) for three different ejector geometries. Simulation data were computed using the $k$-$\omega$ SST turbulence model.}
        \label{fig:valid}
    \end{figure}

    Note that in Figure~\ref{fig:turbms} and later in Figure~\ref{fig:valid}, there are two or more distinct values for each $\dot{Q}_\mathrm{inlet}$. This is due to the fact that all measurements were repeated more times, each time with different total pressure before suction ($\overline{p}^0_{\mathrm{suction}}$). In simulation, $\overline{p}^0_{\mathrm{suction}}$ is an input parameter and was set accordingly to the experimental value. Hence, these series are not differentiated in the aforementioned figures.
    
    Based on comparison with experimental data, the most suitable models seemed to be $k$-$\omega$ SST and $k$-$\varepsilon$. However, simulations that run with the $k$-$\varepsilon$ model had significant problems with stability and convergence speed. Therefore, the $k$-$\omega$ SST model was chosen.
    
    \begin{figure}[htbp]
        \centering
        \includegraphics{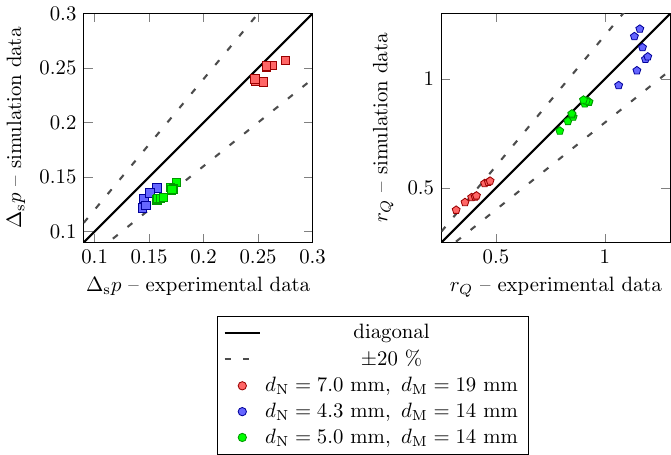}
        \caption{Parity plot showing all experimental and simulation data presented in Figure~\ref{fig:valid} together.}
        \label{fig:parity}
    \end{figure}
     
    \subsection{Model validation}
    \label{sub:validation}
    The prepared CFD model was validated against specifically measured experimental data. The data were measured on three different ejector geometries (specified by $d_\mathrm{N}$ and $d_\mathrm{M}$) and with varying primary fluid flow rates ($\dot{Q}_\mathrm{inlet}$). The simulation and experimental data were compared focusing on the scaled pressure drop of the device ($\Delta_\mathrm{s} p$) and secondary-to-primary flow rate ratio ($r_Q$), defined in Section~\ref{sub:meshSize}.
    
    The comparison of the experimental and simulation data is presented in Figure~\ref{fig:valid} and Figure~\ref{fig:parity}. For a given ejector geometry, the CFD model slightly over- or under-predicts the performance indicators. Nonetheless, the error is always within the range of $\pm 20\ \%$ of its measured value, as shown in Figure~\ref{fig:parity}. Although the agreement in the experimental and simulation data is not perfect, the simulations were able to catch the trends of indicator changes with respect to flow rate and device geometry, see Figure~\ref{fig:valid}. As this simulation capability is essential for geometry optimization, the agreement with the experimental data was deemed acceptable for further use in the optimization framework.

    \setcounter{figure}{0}
    \setcounter{table}{0}
    \section{Additional data from validation of optimization results}
    \label{sec:valprint}
    \begin{figure}[htbp]
        \centering
        \includegraphics{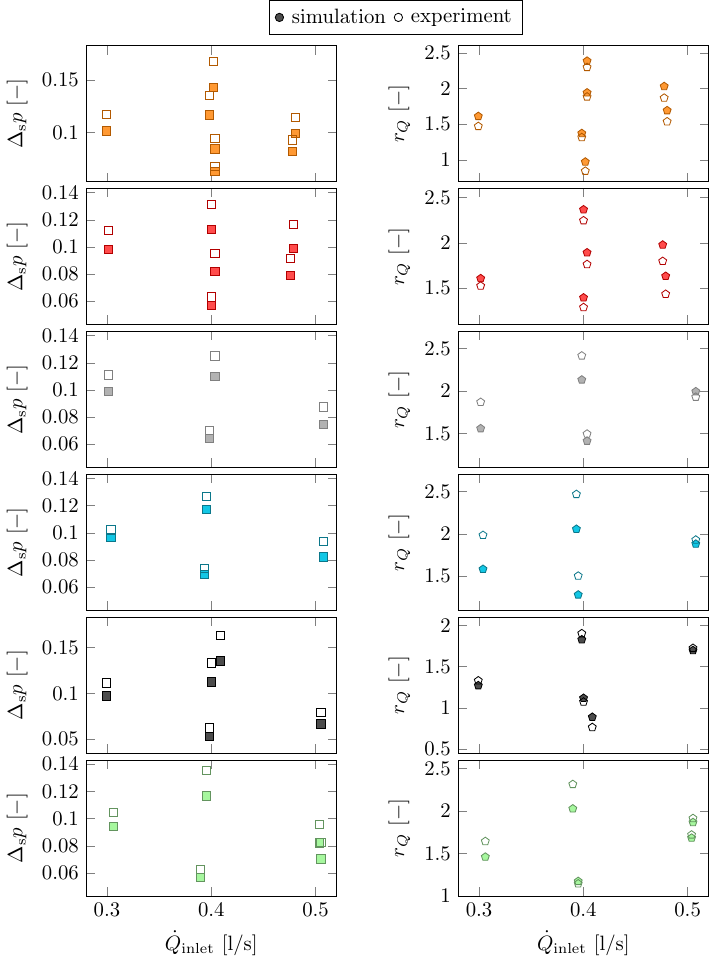}
        \caption{Comparison of simulation and experimental data, namely the scaled pressure drop $(\Delta_\mathrm{s} p)$ and the secondary-to-primary flow rate ratio $(r_Q)$ are depicted for different primary fluid flow rates $\dot{Q}_\mathrm{inlet}$. Colors correspond to ejector designs displayed in Figure~\ref{fig:exppof}.}
        \label{fig:expvscfd}
    \end{figure}
    
    \begin{figure}[htbp]
        \centering
        \includegraphics{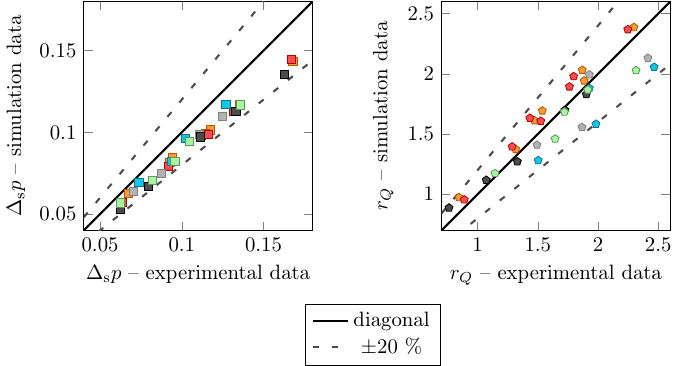}
        \caption{Parity plot showing all experimental and simulation data presented in Figure~\ref{fig:expvscfd} together. Colors correspond to designs displayed in Figure~\ref{fig:exppof}.}
        \label{fig:paropt}
    \end{figure}
    
    The comparison of all available experimental data with the data from the corresponding CFD models is shown in Figures~\ref{fig:expvscfd} and~\ref{fig:paropt}. The comparison is based on two indicators presented in~\ref{sub:validation}, namely the scaled device pressure drop ($\Delta_\mathrm{s} p$) defined in \eqref{eq:scaledDPApp} and the secondary-to-primary flow rate ratio ($r_Q$) defined in \eqref{eq:flowratio}.
    
    The data show trends similar to the data presented in model validation in~\ref{sub:validation}. For a given design, the CFD model tends to over- or under-predict the experimental data, but the error stays in the bounds of $\pm 20\ \%$ of the measured value.
    
    \clearpage
    \setcounter{figure}{0}
    \setcounter{table}{0}
    \section{\revs{Stochastic behavior and architectural preferences of CFDNNetAdapt}}
    \label{sec:heuristics}
    \revs{During each run, CFDNNetAdapt adaptively determines a suitable architecture of the multilayer perceptron (MLP) used as the surrogate model. The procedure of the adaptive MLP architecture selection is described in detail in Section~\ref{sec:cfdnnetadapt} and is based on a random selection of the number of neurons in each hidden layer.}
    
    \revs{In the following, we provide data to examine the variability in the MLP architectures identified by CFDNNetAdapt. Each of the standardized benchmark problems, i.e., ZDT and LZ functions, as well as the final real-life optimization problem were solved repeatedly. For each run, the algorithm was started from the same initial values listed in Table~\ref{tab:setzdts} for ZDT functions, in Table~\ref{tab:setreal} for the real-life application, and in Table~4 in the Supplementary material for LZ functions.}
    
    \begin{table}[htbp]
        \centering
        \small
        \revs{
        \begin{tabular}{cccc}
            \toprule
            function & mean $\bar{n}_1$ (sdev/skew) & mean $\bar{n}_2$ (sdev/skew) & mean $\bar{n}_3$ (sdev/skew) \\
            \midrule
            ZDT1 & 10 (6/+0.4) & 9 (5/+0.6) & 7 (5/+1.0) \\
            ZDT2 & 5 (2/-0.3) & 8 (4/+0.3) & 7 (3/-0.3) \\
            ZDT3 & 8 (4/+0.5) & 9 (4/-0.1) & 12 (4/-0.3) \\
            ZDT4 & 5 (2/-0.3) & 10 (5/-0.2) & 7 (3/0.3) \\
            ZDT6 & 7 (3/+0.2) & 10 (4/+0.7) & 8 (4/+0.6) \\
            \bottomrule
        \end{tabular}
        }
        \caption{\revs{Mean sizes of each hidden layer of the surrogate MLP proposed by CFDNNetAdapt for individual ZDT functions. The means are computed over 15 algorithm runs and the data are complemented by standard deviation (sdev) and skewness (skew) of the proposed hidden layer size.}}
        \label{tab:zdtsarchstats}
    \end{table}
    \revs{In Table~\ref{tab:zdtsarchstats}, the mean sizes of the individual hidden layers to which CFDNNetAdapt converged for ZDT functions are given together with the standard deviation and skewness of these sizes. The mean hidden layer sizes suggest that for each problem, CFDNNetAdapt always preferred specific layer size ratio. However, the heuristic nature of CFDNNetAdapt becomes apparent when comparing the standard deviations of hidden layer sizes to the standard deviations in IGD and $\Delta$HV listed in Table~\ref{tab:zdtsstats}. In particular, when CFDNNetAdapt was able to converge to a good $\pFront$ and $\pSet$ estimates, it was able to do so even with relatively different MLP architectures.}

    \begin{figure}[htbp]
        \centering
        \includegraphics{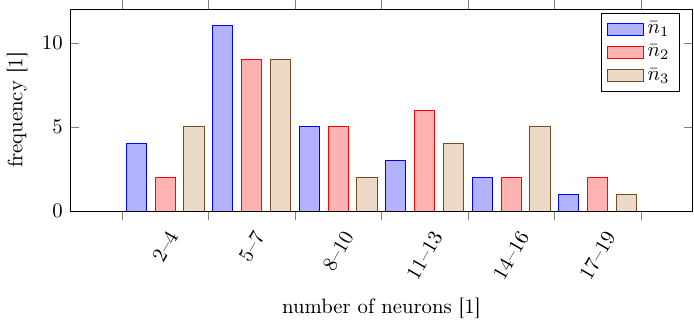}
        \caption{\revs{Histogram for the number of neurons in each hidden layer of the MLPs proposed by CFDNNetAdapt for optimization of the ejector shape. Data for 25 algorithm runs are shown.}}
        \label{fig:histogram_real}
    \end{figure}
    \revs{For the real-life case, 25 individual CFDNNetAdapt test runs were conducted to assess the stochastic behavior of the algorithm. On average, the final hidden layers contained $8$, $9$, and $9$ neurons, respectively, with a standard deviation of four for each layer. The sample skewness values were, in order, $1.2,$ $0.6$ and $0.2$. To provide additional context, a histogram for the number of neurons in each hidden layer is shown in Figure~\ref{fig:histogram_real}. Based on these histograms and the corresponding descriptive statistics, it can be concluded that CFDNNetAdapt consistently preferred networks with 5-7 neurons in $\bar{n}_1$. For $\bar{n}_2$ and $\bar{n}_3$, the distributions are not strictly unimodal, as both exhibit notable peaks at layer sizes exceeding 10 neurons. Still, clear trends in CFDNNetAdapt behavior can be identified, and it is reasonable to expect that with more repeated runs, the algorithm architectural preferences would become even more pronounced.}

\end{appendix}

\end{document}